\documentclass[journal,10pt,final]{IEEEtran}
\usepackage{cite}
\usepackage{amsmath}
\usepackage{amssymb}
\usepackage{xcolor}
\usepackage{amsthm}
\usepackage{mathtools,cuted}
\usepackage{siunitx}
\usepackage{cuted}
\usepackage{graphicx}
\usepackage{subcaption}
\usepackage{epstopdf}
\usepackage{soul}
\usepackage{enumitem}
\usepackage{tabularx}
\usepackage{multirow}
\usepackage{booktabs}
\usepackage{diagbox}

\usepackage{optidef}

\RenewEnviron{BaseMiniExclam}[7]{%
  \selectConstraintMult{#1}%
  \renewcommand{\localOptimalVariable}{#2}%
  \begin{subequations}
  \ifthenelse{\equal{#7}{b}}{\allowdisplaybreaks}%
  #4
  \begin{alignat}{5}
    \bodyobj{#2}{#3}{#6}{#5}
    \BODY
  \end{alignat}
  \end{subequations}%
  \setStandardMini
}

\usepackage[boxruled,norelsize]{algorithm2e}
\makeatletter
\newcommand{\removelatexerror} {\let\@latex@error\@gobble}
\makeatother

\usepackage{tcolorbox}
\tcbuselibrary{many}

\newlength{\flexwidth}
\newcommand{\revisioncolor}{blue}
\renewcommand{\revisioncolor}{black}
\newcommand{\revise}[1]{{\color{\revisioncolor}#1}}

\theoremstyle{plain}
\newtheorem{theorem}{Theorem}
\newtheorem{lemma}{Lemma}
\newtheorem{corollary}{Corollary}
\newtheorem{remark}{Remark}

\newcommand{\superscript}[1]{^{\text{#1}}}
\newcommand{\subscript}[1]{_{\text{#1}}}
\newcommand{\transpose}{\superscript{T}}
\newcommand{\diff}{\text{d}}

\newtcbtheorem[]{alg}{Algorithm}{fonttitle=\bfseries}{alg}

\IEEEoverridecommandlockouts

\begin{document}

\title{
	Fairness for Freshness:\\Optimal Age of Information Based OFDMA Scheduling with Minimal Knowledge
}

\author{Bin~Han,~\IEEEmembership{Senior Member,~IEEE,}
        Yao~Zhu,~\IEEEmembership{Student~Member,~IEEE,}
        Zhiyuan~Jiang,~\IEEEmembership{Member,~IEEE,}\\%
        Muxia~Sun,
        and~Hans~D.~Schotten,~\IEEEmembership{Member,~IEEE}
\thanks{B. Han and H. D. Schotten are with the Division of Wireless Communications and Radio Positioning (WiCoN), University of Kaiserslautern, 67663 Kaiserslautern, Germany. E-mails: \{binhan,schotten\}@eit.uni-kl.de. Y. Zhu is with the research group ISEK, RWTH Aachen University, 52072 Aachen, Germany. Z. Jiang is with Shanghai Institute for Advanced Communication and Data Science, Shanghai University, Shanghai 200444, China. E-mail: jiangzhiyuan@shu.edu.cn. M. Sun is with the Department of Industrial Engineering, Tsinghua University, Beijing 100084, China. E-mail: muxiasun@mail.tsinghua.edu.cn. This work was supported by the German Federal Ministry of	Education and Research (BMBF) research project 5Gang (No. 16KIS0725K), the National Key R$\&$D Program of China (No. 2019YFE0196600), the program for Professor of Special Appointment (Eastern Scholar) at Shanghai Institutions of Higher Learning. The corresponding author is Zhiyuan Jiang.}
}

    \maketitle

\begin{abstract}
It is becoming increasingly clear that an important task for wireless networks is to minimize the age of information (AoI), i.e., the timeliness of information delivery. While mainstream approaches generally rely on the real-time observation of user AoI and channel state, there has been little attention to solve the problem in a complete (or partial) absence of such knowledge. In this article, we present a novel study to address the optimal blind radio resource scheduling problem in orthogonal frequency division multiplexing access (OFDMA) systems towards minimizing long-term average AoI, which is proven to be the composition of time-domain-fair clustered round-robin and frequency-domain-fair intra-cluster sub-carrier assignment. Heuristic solutions that are near-optimal as shown by simulation results are also proposed to effectively improve the performance upon presence of various degrees of extra knowledge, e.g., channel state and AoI.
\end{abstract}

\begin{IEEEkeywords}
Age of information, finite blocklength, OFDMA, radio resource allocation, blind scheduling
\end{IEEEkeywords}

\IEEEpeerreviewmaketitle


\section{Introduction}\label{sec:introduction}
The research attention to timely information delivery in wireless networks has been dramatically increasing over the past few years, due to the emerging demand for time-critical applications such as factory automation and autonomous driving. Recently, the age of information (AoI) concept \cite{kaul2012real} has been intensively investigated as a practical metric for information timeliness. Latest studies have demonstrated that, in many wireless applications such as cooperative autonomous driving~\cite{jiang2020ai}, AoI optimization is more efficient than simple latency minimization.

The scheduling problem for AoI optimizations in wireless networks has been studied in some previous works, usually assuming full knowledge of the channel, i.e., transmission error probability, and global AoI of users \cite{kadota18,maatouk20,tang20,he17,lu18,joo18,abd20}, whereas relaxing the availability of either knowledge leads to a new study direction. Assuming no global AoI information, decentralized scheduling has been investigated \cite{jiang19_tcom,roy18_isit}, showing that it is possible to achieve similar performance as centralized scheduling, when the random access process is properly designed. Few works have considered the vacancy of channel knowledge in AoI optimizations and its impact on both the algorithm design and system performance. A reinforcement learning (RL) approach is proposed in \revise{\cite{gunduz18}} to gradually learn the channel knowledge through trial-and-error. It is also shown that the well-known round-robin (RR) scheduling policy is optimal when the scheduling intervals are independent, i.e., renewal, when the channel is error-free \cite{jiang19_iotj}. On the other hand, the scheduling problem in mainstream OFDM-based systems, wherein there are multiple concurrent channels with adjustable capacity (number of allocated sub-carriers), has not been addressed.

Besides, the aforementioned works are conducted under the infinite blocklength (IBL) regime, where the error rate can be arbitrarily low, if the transmission rate is below Shannon capacity and the blocklength is sufficient. Unfortunately, the assumption of IBL can be usually violated in practical systems. Especially in Internet-of-Things (IoT) networks and ultra-reliable low-latency communication (URLLC) applications, the blocklength can be quite short due to the stringent constraints on latency and radio resource, so that the impact of finite blocklength (FBL) codes shall be incorporated into the analysis~\cite{polyanskiy2010channel}. 
Thanks to the recent advances in characterization of coding rates~\cite{polyanskiy2009dispersion,polyanskiy2015AWGN}, system performance in the FBL regime has been extensively studied. The authors of~\cite{She2019URLLC} proposed a resource allocation scheme to fulfill the requirement of URLLC. The work in~\cite{Hu2016URLLC} analyzed the performance of relay system with FBL codes under fading channels. An optimal framework design for wireless power transfer within FBL regime is provided in~\cite{Hu2019URLLC}. The authors of~\cite{Makki2019HARQ} studied the performance of re-transmission with FBL codes and proposed a fast hybrid automatic repeat request protocols. {While most existing studies on FBL focus to minimize the transmission error probability, there have been some pioneering work reported in the recent past, which investigate the AoI over queues and lossy channels in the FBL regime. For instance, regarding the schemes of channel coding and automatic repeat request (ARQ) in a single-user system, the probabilities of violation in latency and AoI have been analyzed in \cite{devassy2018delay}, and age-optimal designs have been proposed in \cite{sac2018age, devassy2019reliable}. For multi-user FBL systems, with our previous works \cite{han2019optimal,han2020recursive} we have studied the AoI optimization in simple TDMA uplink transmission. Generally, the problem of optimal dynamic selection of blocklength can be modeled as a Markov decision Process (MDP), where the AoI is stochastically determined by both the previous AoI and the channel state in a joint. 
RL-based methods are known to be effective in solving MDP optimization problems, but they essentially require online observations on the (at least partial) global information of both AoI and channel state. Therefore, they are not applicable in absence of such knowledge.}

All the aforementioned approaches can be critically challenged by a realistic use scenario of modern wireless systems, where numerous user equipment (UE) are networked in frequency-selective radio environments, but only a minimal knowledge of global AoI and channel state is available for the scheduler, so that blind scheduling has to be relied on. Such a vacancy of knowledge can be practically common due to a two-folded reason. First, to enable the observation and update of system/channel state, it generally requires to implement several extra control-plane functionalities, e.g. spectrum sensing, logic link control (LLC) layer acknowledgment, etc. These functionalities will significantly increase the system implementation cost and the UE power consumption, which are critical drawbacks for applications requiring large-scale deployment, e.g. smart city and Internet-of-Everything (IoE). Secondly, the observation and update will also generate a significant signaling overhead and an extra delay, which is a serious issue in timeliness critical applications, such as factory automation and autonomous driving. 

Among typical blind schedules, RR has been commonly used in existing studies on AoI \cite{jiang19_iotj} as a baseline, for its simple implementation and predictable performance. It can be described as always selecting the UE to transmit who has the highest interval since its last transmission (no matter if successful). Especially, for error-free channels, RR is equivalent to the AoI-greedy schedule, which is the global optimum among all causal schedules \cite{kadota18}. However, when it comes to lossy channels, the superiority of RR over other blind schedules is not as axiomatic as it may intuitively appear. For example, when the packet error rate (PER) of every single transmission attempt is very high, it is guaranteed to effectively reduce the short-term AoI of the UE in turn to transmit, by assigning two slots in a row to it. This will, on the other hand, increase the short-term AoI for all other UEs. However, how such short-term effects will converge to a long-term impact, upon different channel conditions and various AoI status of UEs, is not straightforward.

To address this issue, in this work we propose an AoI optimal blind FBL-OFDMA scheduler with minimal requirement of knowledge about system and channels. \revise{Compared to literature, the research presented in this manuscript exhibits its main novelty in two aspects: \emph{i}) Most existing AoI-optimal approaches, such as \cite{gunduz18,han2019optimal}, and \cite{han2020recursive}, rely on a rich knowledge of instantaneous system status, including different UEs' AoI, channel state information (CSI), PER, and the LLC layer feedback of every transmission attempt. Our method, to the best of our knowledge, is the first proposal to deal with AoI-optimal scheduling in lack of real-time system status information. \emph{ii})  While the mainstream studies on AoI optimization, such as \cite{kadota18,jiang19_tcom,gunduz18}, and \cite{jiang19_iotj}, consider the TDMA scheduling problem where only one UE is granted transmission opportunity in every slot, here we investigate the scheduling problem in OFDMA systems, where multiple UEs can share the sub-carriers in a same time slot. It therefore introduces into discussion the new challenges of intra-cluster sub-carrier assignment and cluster sizing.}

\revise{More specifically, \emph{i}) In the time domain, we prove that as long as the PER is independent and identically distributed (i.i.d.) for all UEs over all time slots, the clustered RR policy becomes an asymptotic optimal blind schedule that minimizes the average AoI as the time horizon approaches towards infinite, regardless of the average PER and the initial AoI status. This theorem, which is yet never proven by literature, identifies the AoI-optimal schedule in condition of minimal knowledge. \emph{ii}) In the frequency domain, we prove that as long as the signal-to-noise ratio (SNR) is i.i.d. for all UEs among all sub-carriers, a random uniform allocator becomes the AoI optimal solution of intra-cluster sub-carrier assignment in lack of accurate CSI. This design does not only resolve the intra-cluster sub-carrier assignment problem, but also guarantees the PER heterogeneity required by the clustered RR scheduler. Furthermore, upon the long-term average PER among all UEs, we are able to solve the optimal cluster size for the proposed scheduler, which is \emph{independent} of the total amount of UEs. Thus, the cluster sizing issue is also addressed. \emph{iii}) Additionally, to enhance the applicability of our proposed approach in more generic scenarios, where the scheduler may be able to obtain additional knowledge, e.g. LLC-layer acknowledgment and UE-specific CSI, we provide a three-stage framework on top of the blind clustered RR scheduler, which allows the system to leverage such information, so as to further heuristically improve the AoI performance to near-optimum shown by simulation results.}

The remainder of the paper is organized as follows. In Sec.~\ref{sec:model} we present the scenario of clustered radio resource scheduling in OFMDA, setup the system model and the AoI optimization problem. In Sec.~\ref{sec:analysis} we analyze this problem in the minimal knowledge case, and propose the optimal blind scheduler, which is a clustered round-robin with optimized cluster size, combined with a random uniform intra-cluster sub-carrier assignment. To cope with a more generic case where advanced knowledge may be available, in Sec.~\ref{sec:heuristics} we propose a three-stage optimizing framework that invokes two heuristic methods on the top of our proposed optimal blind scheduler, which exploit the extra knowledge for further AoI reduction. All proposed methods are numerically evaluated in Sec.~\ref{sec:simulations}, before we close the article with a conclusion. The most important notations used in this article are collectively listed with explanations in Tab.~\ref{tab:notations}.

\begin{table*}[!htbp]
	\centering
	\caption{Important notations}
	\label{tab:notations}
\begin{tabular}{|l|l|c|l|l|}
		\cmidrule[2px]{1-2}\cmidrule[2px]{4-5}
		\textbf{Notation}	&\textbf{Meaning}									&&\textbf{Notation}			&\textbf{Meaning}\\
		\cmidrule[1px]{1-2}\cmidrule[1px]{4-5}
		$\mathcal{I}$ 		&set of all UEs $\{1,2,\dots,I\}$					&&$r_i$ 					&coding rate of UE $i$\\
		$\mathcal{M}$ 		&set of all sub-carriers $\{1,2,\dots,M\}$			&&$n_{m,i}$ 				&time of UE $i$ occupying sub-carrier $m$ per transmission\\
		$\mathcal{M}_i$ 	&set of sub-carriers assigned to UE $i$				&&$a_{\pi,m,i}\revise{(t)}$ &indicator: sub-carrier $m$ assigned to UE $i$ \revise{in time slot $t$}\\
		$l$ 				&UE cluster size									&&							&under schedule $\pi$\\
		$\tau$ 				&information bits per packet						&&$\gamma_{m,i}$			&SNR for UE $i$ on sub-carrier $m$\\
		$\varepsilon_i$ 	&packet error rate of UE $i$						&&$C_{m,i}$ 				&normalized channel capacity of sub-carrier $m$ for UE $i$\\
		$\lambda_i$ 		&\revise{probability of new packet generation}		&&$V_{m,i}$ 				&normalized channel dispersion of sub-carrier $m$ for UE $i$\\
							&\revise{per slot by UE $i$}						&&							&\\
		\cmidrule[2px]{1-2}\cmidrule[2px]{4-5}
	\end{tabular}
\end{table*}

\section{System Model}\label{sec:model}
\begin{figure}[!h]
	\centering
	\includegraphics[width=0.9\linewidth]{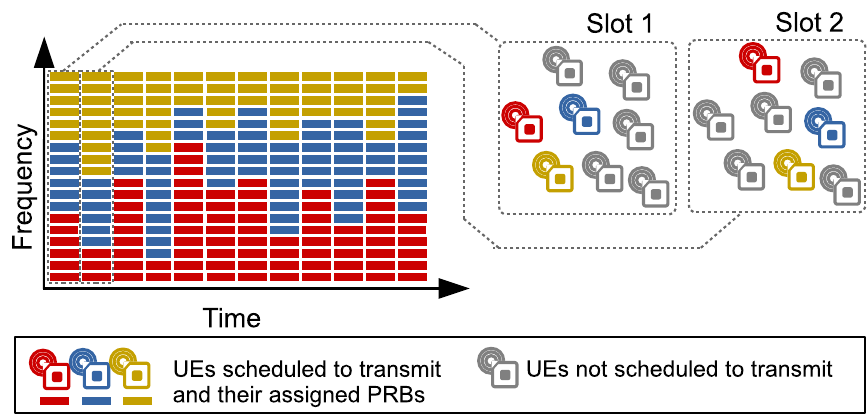}
	\caption{A minimal example of the proposed PRB scheduling scheme where $I=8$ and $l=3$}
	\label{fig:mechanism}
\end{figure}

We consider a use scenario with massive deployment of low-cost, power-efficient UEs, e.g. a massive IoT network. All UEs in the system, notated as the set $\mathcal{I}$, are registered at the multi-access edge computing (MEC) server. Like most mainstream solutions, we consider in our model a synchronized OFDMA system, where the radio resource is divided into orthogonal physical resource blocks (PRBs). Since the number of UEs $I\triangleq\vert\mathcal{I}\vert$ can be huge in the discussed scenario, the amount of sub-carriers is usually inadequate to support simultaneous uplink transmissions of all UEs, therefore the sub-carriers must be scheduled in a slotted manner. The time domain is equally divided into transmission slots of length $n$ (\emph{note that for convenience of notation, all time values in this article, including AoI, are normalized to the symbol length}), which is shorter than the channel coherence time. In every transmission slot, $l$ UEs are selected from the UE set $\mathcal{I}$ as the \emph{active cluster}, which share the set $\mathcal{M}$ of $M$ sub-carriers to individually transmit messages in uplink to the MEC server. We consider all messages transmitted by all UEs to have the same payload information bits, and to be encoded with the same channel coder under the same coding rate. This mechanism is briefly illustrated by Fig.~\ref{fig:mechanism}. 

Both the selection of active cluster, and the sub-carrier assignment within each cluster, shall be carried out w.r.t. some consistent schedule. Such a schedule $\pi$ can be uniquely addressed by a resource assignment sequence $\{\mathbf{A}_\pi(t)\vert t=1,2,\dots\}$, where for arbitrary slot index $t$, every element $a_{\pi,m,i}(t)$ in the resource assignment matrix $\underset{M\times I}{\mathbf{A}_\pi}(t)$ is a PRB assignment index that:
\begin{equation}
	a_{\pi,m,i}(t)=\begin{cases}
		1&\text{sub-carrier $m$ assigned to UE $i$ in slot $t$,}\\
		0 &\text{otherwise.}
	\end{cases}
\end{equation}
Therewith, we can define the active cluster in time slot $t$ as $\mathcal{I}_t\triangleq\left\{i:i\in\mathcal{I},\sum\limits_{m\in\mathcal{M}}a_{\pi,m,i}(t)\geqslant 1\right\}$. Especially, since online schedulers that rely on up-to-date system status require significant implementation complexity, power consumption, and signaling overhead, as we have discussed in Sec.~\ref{sec:introduction}, we are interested in those \emph{blind} schedules that require only minimal system status information, namely the previous schedule history and the long-term global average SNR over all UEs and sub-carriers. That means, for all $t\in\mathbb{N}^+$:
\begin{equation}
	\mathbf{A}_\pi(t)=f\left[\mathbf{A}_\pi(1),\mathbf{A}_\pi(2),\dots,\mathbf{A}_\pi(t-1),\bar{\gamma}\right],\label{eq:blindness}
\end{equation}
where $\revise{\bar{\gamma}=}\lim\limits_{T\to+\infty}\frac{1}{TMI}\sum\limits_{t=1}^T\sum\limits_{m=1}^M\sum\limits_{i=1}^I\gamma_{m,i}(t)$.

Now we investigate the AoI of different UEs in such a system. At the end of any slot $t$, the AoI of an arbitrary UE $i\in\mathcal{I}$ at the server is $h_i(t)=nt-u_i(t)$, where $u_i(t)$ is the birth time of the last message that $i$ has successfully transmitted to the server until $t$. Without loss of generality, we consider all UEs as active sources, i.e. a UE always generates a fresh message \revise{every slot. Furthermore, here we take advantage of the well-known principle in AoI-sensitive system design, that a Last-Come-First-Serve (LCFS) queuing policy with queue length of 1 is optimal for the buffer of message sources~\cite{yates2019age}, i.e. the old messages shall always be simply dropped when a new message arrives. In this way, the message update model can be equivalently considered as every UE generating a new message when it is scheduled to transmit\footnote{\revise{Remark that this M/M/1 LCFS queuing policy denies any kind of ARQ/HARQ mechanism across multiple transmission slots when assuming active sources.}}}. Thus, with noisy channels, at the end of slot $t\in\mathbb{N}^+$, the AoI of UE $i\in\mathcal{I}$ is completely determined by the successes/failures of its transmission attempts:
\begin{equation}
	\label{eq:AoI}
	h_i(t)=\begin{cases}
		n&s_i(t)=1,\\
		h_i(t-1)+n&\text{otherwise}.
	\end{cases}
\end{equation}
Here, $s_i(t)\sim\mathfrak{B}\left(1-\varepsilon_i(t)\right)$ is a Bernoulli process indicating successful transmissions, where $\varepsilon_i(t)$ is the uplink transmission PER of UE $i$ in slot $t$, i.e. $\varepsilon_i(t)=1$ if $i$ is not scheduled to transmit in slot $t$.
Our aim is to find from the space of all blind schedules $\Pi$ an optimum $\pi$, which minimizes the expected average AoI\footnote{\revise{It is found that time-average AoI is a quantity that combines time-average peak-AoI and its second moment (see \cite{kadota18} for details). Therefore, we argue that minimizing time-average AoI is in fact minimizing both the average peak-AoI and its variance, thus reducing the probability that peak-AoI exceeds a threshold, by invoking the Chebyshev's inequality.}} of all UEs on the infinite horizon:
\begin{mini!}[2]
	{\pi\in\Pi, l\in\mathcal{I}}{\overline{h}_\pi^{(\infty,I)}}{\label{prob:ofdm_scheduling}}{}
	\addConstraint{\vert\mathcal{I}_t\vert=l,\quad\forall t\in\mathbb{N}}
	\addConstraint{\revise{\sum\limits_{i\in \mathcal{I}_t} a_{\pi,m,i}(t)= 1, \forall (m,t) \in \mathcal{M}\times\mathbb{N}}}
	\label{con:n_k}
	\addConstraint{\revise{\varepsilon_i(t){{\leqslant \varepsilon_{\max},  \forall i \in \mathcal{I}_t }}}}
	\label{con:Shannon}
\end{mini!} where $\overline{h}_\pi^{(\infty,I)}\triangleq\lim\limits_{T\to\infty}\overline{h}_\pi^{(T,I)}\triangleq\mathbb{E}_{\pi}\left(\lim\limits_{T\to+\infty}\frac{1}{IT}\sum\limits_{t=1}^T\sum\limits_{i\in\mathcal{I}}h_i(t)\right)$. 
Remark that the constraint~\eqref{con:n_k} ensures that all sub-carriers are assigned, and the constraint~\eqref{con:Shannon} prevents waste of radio resource with a transmission-rejecting PER threshold of $\varepsilon_{\max}=0.5$.


\section{Analyses: AoI-Optimal Blind Scheduling}\label{sec:analysis}
The optimal scheduling problem \eqref{prob:ofdm_scheduling}, as an two-dimensional integer programming task with strong constraints, is hard to solve. Nevertheless, in context of blind scheduling, it can be decomposed, as we will do below in this section, into three sub-stages, namely \emph{i}) the scheduling of active clusters $\mathcal{I}_t$, \emph{ii}) the intra-cluster assignment of sub-carriers, and \emph{iii}) the optimal $l$ sizing of active clusters. We will show that, these three problems can be independently solved, and their optimums jointly support each other to assemble a solution of \eqref{prob:ofdm_scheduling}.

\subsection{Blind Scheduling with Minimal Knowledge: Clustered Round-Robin}\label{subsec:rr_optimum}
First we consider a scheduling of active clusters $\mathcal{I}_t$, when the cluster size $l$ and the sub-carrier assignment policy are fixed. In any time slot $t$, we consider to possess only the full history of previous active cluster scheduling $\mathcal{I}_1,\mathcal{I}_2,\dots,\mathcal{I}_{t-1}$, as well as the long-term PER averaged over space and time among all UEs \revise{$\varepsilon\subscript{\revise{avg}}\triangleq\lim\limits_{T\to+\infty}\frac{1}{TI}\sum\limits_{t=1}^T\sum\limits_{i\in\mathcal{I}}\varepsilon_i(t)$}, %
and look for an optimal schedule that minimizes the expected average AoI. Starting with the special case $l=1$, we obtain that:
\begin{theorem}\label{th:blind_optimum_rr}
	  \revise{When $l=1$, f}or homogeneous UEs \revise{that have independent and identically distributed time-varying PERs}, among all blind schedules, round-robin is (1) the optimum that minimizes $\frac{1}{I}\sum\limits_{i\in\mathcal{I}}\mathbb{E}_\pi\left(h_i(t)\right)$ for all $t$, if $h_i(0)=h_j(0) \forall i,j\in\mathcal{I}$; and (2) an asymptotic optimum that, with arbitrary initial AoI $h_i(0)$ for all $i\in\mathcal{I}$, minimizes $\overline{h}_\pi^{(T,I)}$ as $T\to+\infty$.
\end{theorem}
This theorem can be proven as follows (a detailed proof by us is provided in Appendix~\ref{app:opt_rr_proof}):
\begin{enumerate}
	\item Without loss of generality, starting from a zero initial status where all UEs have AoI of 1, the expected average AoI can be modeled as a convex decreasing utility function over a special probability space defined with the inter-transmission interval (ITI).
	\item RR second-order stochastically dominates any other deterministic schedule, so that it is the optimal blind schedule for zero initial status that minimizes the expected average AoI.
	\item The impact of initial status on the expected average AoI vanishes on the infinite horizon, so that RR becomes an asymptotic optimum for arbitrary initial status.
\end{enumerate}

\revise{Theorem \ref{th:blind_optimum_rr} can be easily generalized for any $l\in\mathcal{I}$ with an $l$-clustered round-robin schedule, which transmits in every slot the $l$ UEs with the longest interval since their last transmission (see proof in Appendix~\ref{app:opt_rr_proof}):}
\revise{\begin{corollary}\label{cr:blind_optimum_crr}
	  For arbitrary $l\in\mathcal{I}$ and homogeneous UEs that have independent and identically distributed time-varying PERs, among all blind schedules, $l$-clustered round-robin is an asymptotic optimum that, with arbitrary initial AoI $h_i(0)$ for all $i\in\mathcal{I}$, minimizes $\overline{h}_\pi^{(T,I)}$ as $T\to+\infty$.
\end{corollary}}

Now we further analyze the AoI performance of RR policy. We know that given the UE-specific transmission error probability $\varepsilon_i$, and the UE-specific \revise{probability $\lambda_i$ of generating new packet per slot} for all $i\in\mathcal{I}$, the average AoI of UE $i$ under RR on the infinite horizon \revise{is}
\begin{IEEEeqnarray}{rCl}\label{eq:cons_cond}
	\bar{h}_{i,\text{RR}}^{(\infty,I)}  =  \frac{1}{\lambda_i} + \frac{n\left(1+\varepsilon_i\right)}{2(1-\varepsilon_i)}\alpha +n  -\frac{3}{2}, 
\end{IEEEeqnarray}
where $\alpha \triangleq I/l$ denotes the ratio between the total number of users and the number of scheduled users in each time slot, and hence $\alpha$ also denotes the scheduling interval of each user (in terms of OFDM symbols). The proof of \eqref{eq:cons_cond} is based on a generalized arrival theorem of Poisson process and the details can be found in our previous work in \cite[Theorem 1]{jiang18GC}.

In our system, only the long-term average PER $\varepsilon\subscript{\revise{avg}}$ is known at the scheduler, and all UEs are active sources, i.e. $\lambda_i=1,\forall i\in\mathcal{I}$. Thus, the average AoI among all users is
\begin{equation}
    \bar{h}\subscript{RR}^{(\infty,I)} =  \frac{1}{I} \sum\limits_{i\in\mathcal{I}}\left[1 +  \frac{n(1+\varepsilon\subscript{\revise{avg}})}{2(1-\varepsilon\subscript{\revise{avg}})}\alpha  +n  -\frac{3}{2}\right] .
    \label{eq:avg_aoi}
\end{equation}

\revise{Before leaving this cluster scheduling stage and heading to sub-carrier assignment, we consider it worth to review the assumption of i.i.d. PER that we take in this subsection. Generally, it is common and practical to consider i.i.d. SNR among UEs, especially in those use scenarios where UEs are homogeneous in device type and mobility pattern. As we will see in the next subsection, when no CSI is available, our blind sub-carrier assignment solution tends to randomly and uniformly allocate all available sub-carriers to different users within every cluster $\mathcal{I}_t$. With i.i.d. SNRs among all UEs and all time slots, it obviously leads to i.i.d. PERs $\varepsilon_i(t)$ for all $(i,t)$.

Yet we shall note, that heterogeneity and inter-user coherence in channel conditions are also commonly observed in some real world use scenarios. For example, when users move in a clustered pattern (which is typical in indoor scenarios and vehicle platoons, leading to a SNR coherence among different UEs), when UEs are of heterogeneous device types (where the antenna gain and mobility model varies from user to user), or when the UEs are fixed in location (where the spatial selectivity in path loss and shadowing effect is no more balanced over time by the user mobility). In such cases, the blind sub-carrier assignment cannot guarantee i.i.d. PERs. Thus, when scheduling the transmission opportunities, it will be naturally better to discriminate against the UEs that are more likely to have lower PERs. Nevertheless, for the \emph{blind} scheduler under our study, which possesses no CSI other than $\varepsilon\subscript{avg}$, it is impossible to implement such a discriminative policy. Without specific knowledge about the heterogeneous PERs, an arbitrary discrimination against randomly picked UEs can only be expected to worsen the performance. In fact, even only to be aware of any heterogeneity or inter-user coherence in channel conditions, it inevitably requires some CSI. In the point of view by a scheduler with minimal knowledge, the PERs of different UEs, despite of their heterogeneity or coherence, still exhibit no characteristic else than an independent and identical distribution. 

Moreover, when the SNRs are not i.i.d. but can be obtained, we can make use of this CSI on the sub-carrier assignment sub-stage for an improved performance, as we will propose in Sec.~\ref{subsec:recursive_subcarrier_assignment}. Remarkably, such CSI-based sub-carrier assignment still always yells for balancing the PER among all UEs to the same level. The reason is that the condition of optimal sub-carrier assignment, provided by Eq.~\eqref{eq:eps_uniform} in Appendix~\ref{app:blind_assignment}, holds regardless of knowledge level on the CSI. Thus, a universal PER for all UEs is achieved, which is an extreme case of i.i.d. PERs, so the applicability of Theorem 1 is still not shaken.
}


\subsection{Blind Sub-carrier Assignment with Minimal Knowledge: Uniform Assignment}\label{subsec:subcarrier_assignment}



{Now we turn to the intra-cluster sub-carrier assignment problem, where in an arbitrary slot $t$, the active cluster with $l$ UE has been selected, and $M$ sub-carriers shall be allocated to them in order to minimize the cluster-average AoI after the transmission slot. {While the SNR $\gamma_{m,i}$ of sub-carrier $m$ for UE $i$ is random and i.i.d. for all $(m,i)\in\mathcal{M}\times\mathcal{I}_t$}, and the initial AoI status of all UEs are exogenously determined by the selection of active cluster $\mathcal{I}_t$, we consider that {the sub-carrier assignment module only possesses the minimally statistical knowledge about the channel and AoI status, i.e. the SNR distribution and the average AoI.}

Since the blocklength of each message is finite, the transmission of every UE can be erroneous. We start with a generic analysis under such FBL conditions.  In particular, let $\tau$ denote the message length, following the FBL transmission model~\cite{polyanskiy2010channel}, the 
PER of any UE $i \in \mathcal{I}_t$  is given by}
\begin{equation}
\begin{split}
\label{eq:single_link_error_pro}
\varepsilon_{i}\revise{(t)} \approx Q\left(\frac{\left(\sum\limits_{m\in\mathcal{M}}a_{\pi,m,i}\revise{(t)}C_{m,i}n-\tau\right)\ln 2}{\sqrt{\sum\limits_{m\in\mathcal{M}}a_{\pi,m,i}\revise{(t)}V_{m,i}n}}\right),
\end{split}
\end{equation}
 where  ${{C}_{m,i}} = {\log _2}( {1 + \gamma_{m,i} } )$ is the Shannon capacity normalized to \SI{1}{\hertz}, and $V_{m,i}$ is the channel dispersion~\cite{chen2015ubiquitous} of UE $i$ via sub-carrier $m$. Under a complex AWGN channel, $V_{m,i}=1-\frac{1}{(1+\gamma_{m,i})^2}$. {Furthermore, as shown in~\cite{Makki2019HARQ}, the one-shot scheme always outperforms any retransmission scheme with given blocklength $\sum_{m\in\mathcal{M}} a_{\pi,m,i}(t)$. Therefore, the transmission of UE $i$ is carried out with all assigned sub-carriers.}    For any given schedule $\pi$ in a certain slot $\tilde{t}$, the objective function in~\eqref{prob:ofdm_scheduling} can be rewritten as:
 \begin{align}
     &\lim\limits_{T\to+\infty}\mathbb{E}_{\pi}\left(\frac{1}{IT}\sum\limits_{t=1}^T\sum\limits_{i\in\mathcal{I}}h_i(t)\right)\\
     =&\lim\limits_{T\to+\infty}\mathbb{E}_{\pi}\left(\frac{1}{IT}\left(\sum\limits_{t\neq\tilde{t}}^T\sum\limits_{i\in\mathcal{I}}h_i(t)+\sum\limits_{i\not\in\mathcal{I}_t}h_i(\tilde{t})+\sum\limits_{i\in\mathcal{I}_t}h_i(\tilde{t})\right)\right)\nonumber
 \end{align}
 \revise{Recall that the assignment is only carried out in the cluster with selected $l$ UEs}, which has no impact to other UEs outside of the cluster. 
The optimization problem is then formulated as:

\begin{mini!}[2]
	{\mathbf{A_\pi}}{\sum\limits_{i\in \mathcal{I}_t}h_i(t)}
	{\label{problem_blocklength}}{}
	\addConstraint{\sum\limits_{i\in \mathcal{I}_t}a_{\pi,m,i}(t)= 1,~\varepsilon_i(t)\leqslant\varepsilon_{\max}}\nonumber
	\addConstraint{a_{\pi,m,i}(t)\in\{0,1\},\forall (m,i) \in \mathcal{M}\times\mathcal{I}_t}
\end{mini!} 

{Note that it requires full knowledge of instantaneous CSI of all links and AoI of all UEs in $\mathcal{I}_t$ to solve the global optimum, while we study here the blind assignment scheme base on only the statistical knowledge. In particular, let $\mathbf{Z} \in \mathbb{R}^{M \times l}$ denote channel matrix in time slot $t$ and its element $z_{m,i}$ denote the channel gain of UE $i$ via sub-carrier $m$ in time slot $t$.
{ Consider all channels for each UE as i.i.d. under a general distribution,} for which we define the PDF of channel gain as} $f_Z (z_{m,i})=f_{Z_{m,i}}(z_{m,i})$.
Then, \revise{$\varepsilon_i(t)$ in from~\eqref{eq:single_link_error_pro} can be reformulated as}
\begin{align}
\label{eq:single_link_error_int}
\varepsilon_{i}\revise{(t)} =\int &\left[Q\left(\frac{\left(\sum\limits_{m\in\mathcal{M}}a_{\pi,m,i}\revise{(t)}C_{m,i}(z_{m,i})n-\tau\right)\ln 2}{\sqrt{\sum\limits_{m\in\mathcal{M}}a_{\pi,m,i}\revise{(t)}V_{m,i}(z_{m,i})n}}\right)\right.\nonumber\\
&\times\left.\prod\limits_{m\in\mathcal{M}} f_Z(z_{m,i})\right]\diff\mathbf{Z}.
\end{align}
As all channels are i.i.d., the  assignment of any sub-carrier $m\in \mathcal{M}$  has the same impact to any UE $i \in \mathcal{I}_t$, i.e., it holds that $\varepsilon_{i}\revise{(t)}=\varepsilon_{j}\revise{(t)}$ if $\sum\limits_{m \in \mathcal{M}} a_{\pi,m,i}\revise{(t)}=\sum\limits_{m \in \mathcal{M}} a_{\pi,m,j}\revise{(t)}, \forall (i,j)\in \mathcal{I}_t$.
In other words, the blind sub-carrier assignment is equivalent to allocating the blocklength of assigned sub-carriers without accounting the index $m$.
 Therefore, we substitute $A_\pi$ in Problem~\eqref{problem_blocklength} with $\mathbf{b}_\pi$, where $b_{i}\revise{(t)}=\sum\limits_{m \in \mathcal{M}} a_{\pi,m,i}\revise{(t)}\in\mathbb{N}^+$ is the number of assigned sub-carriers for UE $i$ . Furthermore, let $\hat{h}$ \revise{denote} the average AoI of UEs. Then, the expected AoI for UE $i$ can be rewritten as
\begin{equation}
\begin{split}
&\mathbb{E}_\pi\{h_i\}=(\hat{h}+{n)\varepsilon_{i} + n(1-\varepsilon_{i})} \\
=&\hat{h} \int\left[Q\left(\frac{\left(b_{i}\revise{(t)}\sum\limits_{m\in\mathcal{M}}C_{m,i}(z_{m,i})n-\tau\right)\ln 2}{\sqrt{b_{i}\revise{(t)}\sum\limits_{m\in\mathcal{M}}V_{m,i}(z_{m,i})n}}\right)\right.\\
&\left.\times\prod\limits_{m\in\mathcal{M}} f_Z(z_{m,i}) \right]\diff\mathbf{Z}{+n,}
\end{split}
\end{equation}
{Therefore, minimizing the average AoI with minimal knowledge is equivalent to minimizing the sum of PER among the UEs.
Furthermore,since we have only the statistical CSI, constraint~\eqref{con:Shannon} is equivalent to $\sum\limits_{m\in \mathcal{M}}a_{\pi,m,i}\revise{(t)}C_{m,i}(\hat{z})n-\tau=b_{i}\revise{(t)}\sum\limits_{m\in\mathcal{M}}C_{m,i}(\hat{z})n-\tau\geqslant0,  \forall i \in \mathcal{I}_t$ with $\varepsilon_{\max}=0.5$, where $\hat{z}=\mathbb{E}(z)$. It \revise{is} worth to mention that \revise{both the equivalences hold} regardless of \revise{the} channel dispersion $V_i$.}  
Hence, we standardize \eqref{problem_blocklength} into the following integer programming problem:
\begin{mini!}[2]
	{\mathbf{b}_\pi}{\sum\limits_{i\in\mathcal{I}_t}\varepsilon_i}
	{\label{problem_blocklength_blind}}{}
	\addConstraint{\sum\limits_{i\in \mathcal{I}_t}b_{i}(t)=M}
	\addConstraint{b_{i}(t)\sum\limits_{m\in\mathcal{M}}C_{m,i}(\hat{z})n-\tau\geqslant 0,~\forall i\in\mathcal{I}_t}
	\addConstraint{b_{i}\revise{(t)}}{\in \mathbb{N}^{+}},
	\label{con:b_integer}
\end{mini!} 
which can only be solved by inefficient exhaustive searching. To address this, we relax the integer  constraint~\eqref{con:b_integer} into $0 \leqslant b_{i}\revise{(t)} \leqslant M$, $\forall b_{i}\revise{(t)} \in \mathbf{b}_t$. The the relaxed problem can be then  characterized by the following lemma (see detailed proof in Appendix.~\ref{app:blind_assignment}):
\begin{lemma}
    \label{lemma:blind assignment}
    \revise{When $\gamma_{m,i}$ is i.i.d. for all $(m,i)$,} the optimal sub-carrier assignment is to uniformly allocate the sub-carriers to each UEs, i.e., $b_{1}=b_{2}=...=\frac{M}{l}$.
\end{lemma}
\revise{
\begin{remark}
	The results of Lemma 1 are conducted based on minimal knowledge. It can be intuitively interpreted as follows: If no information of individual channels/AoI is available, every UE will be treated as homogeneous node and the sub-carriers are assigned uniformly. However, if we have (partial) information regarding CSI and/or spatial distribution of UEs, Problem~\eqref{problem_blocklength_blind} can still be solved efficiently as a convex problem with the computational complexity of $\mathcal{O}(l^2)$.
\end{remark}		
}

Thus, the optimal sub-carrier assignment can be carried out as follows: {First, each UE $i\in\mathcal{I}_t$ is initially assigned with $\lfloor \frac{M}{l} \rfloor$ sub-carriers, i.e., with coding rate $r=\lceil \frac{\tau l}{M} \rceil$.} Subsequently, we randomly choose $(M\mod l)$ UEs and assign one sub-carrier to each, to fully utilize sub-carriers without selective diversity.

\subsection{Optimal Cluster Size Selection with Minimal Knowledge}\label{subsec:cluster_size_selection}
\begin{figure*}[!htbp]
	\centering
	\includegraphics[width=.8\linewidth]{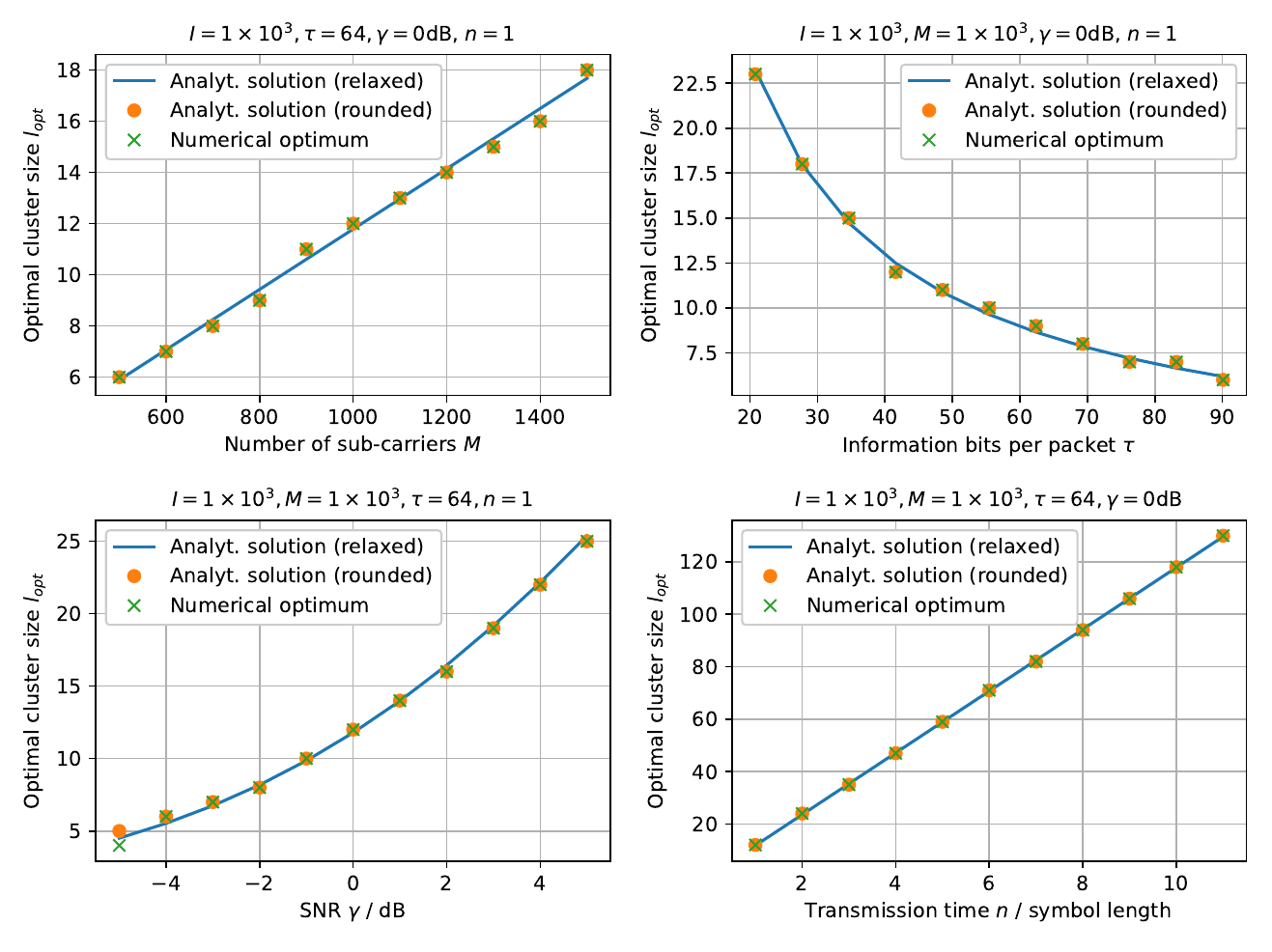}
	\caption{Sensitivity test on the cluster size optimization w.r.t. different system parameters.}
	\label{fig:opt_cluster_size_sensitivity_test}
\end{figure*}

{From the discussions above, we see that the time-domain-fair clustered RR schedule proposed in Sec.~\ref{subsec:rr_optimum} and the frequency-domain-fair random uniform sub-carrier assignment in Sec.~\ref{subsec:subcarrier_assignment} couples well with each other. On the one hand, clustered RR makes an AoI-optimal blind active cluster selection scheme, when all UEs share the same mean PER; in the long term it obviously converges to a situation, that all UEs in every active cluster share the same expectation of AoI. On the other hand, the random uniform assignment is AoI-optimal under blindness, i.e. when assuming all UEs in the cluster to share the same average AoI (which is guaranteed by the clustered RR); and it grants every UE in the active cluster the same PER by means of expectation (which is required by the clustered RR).
Thus, they assemble a complete solution for the two-layer blind PRB scheduling scheme, where the long-term average PER $\varepsilon\subscript{\revise{avg}}$ is used as the PER estimation for all UEs,} i.e. assuming $\bar{h}_{i,\pi}^{(\infty,I)}=\bar{h}_{\pi}^{(\infty,I)},\forall i\in\mathcal{I}$. In such a scenario we can derive the following theorem that identifies the optimal cluster size:
\begin{theorem}\label{th:l-opt}
	\revise{When homogeneous UEs with i.i.d. time-varying PERs are uniformly clustered} for round-robin transmission over a frequency-flat channel, the optimal cluster size (relaxed to real number) that minimizes the long-term average AoI is given by 
	\begin{align}
		l\subscript{opt}&=\frac{1}{2\tau'^2}\times\left(\delta-\sqrt{\delta^2-4\tau'^2M^2C'^2n^2}\right),\label{eq:opt_cluster_size}\\
		\delta&\triangleq -Mn(2C'\tau'+wV),\label{eq:delta}\\
		w&=W_{-1}\left(-2\pi e^{-\frac{4C'\tau'}{V}}\right),\label{eq:lambert_root}
	\end{align}
	where $W_q(\cdot)$ denotes the branch with index $q$ of Lambert $W$ function.
\end{theorem}
The proof of Theorem~\ref{th:l-opt} is provided in Appendix~\ref{app:l-opt}. It is worth to remark that Eq.~\eqref{eq:opt_cluster_size} significantly implies that \emph{with homogeneous UEs and frequency-flat channel, the AoI-minimizing cluster size $l\subscript{opt}$ is independent from the UE amount $I$.}
To verify Theorem~\ref{th:l-opt}, here we conduct a set of numerical tests, where $1000$ UEs are scheduled to transmit in a clustered round-robin fashion. We compare the analytical solution provided by \eqref{eq:opt_cluster_size} and the exhaustively searched optimum that minimizes the $\bar{h}_\pi^{(\infty,I)}$ in \eqref{eq:avg_aoi}, the results are plotted in Fig.~\ref{fig:opt_cluster_size_sensitivity_test}, showing a perfect match between our analysis and the numerical optimum.

\section{Extension: Heuristic AoI Enhancement upon Extra Knowledge}\label{sec:heuristics}

{So far we have proposed an AoI-optimal blind solution for clustered OFDMA scheduling where only minimal knowledge is available at the scheduler. Nevertheless, in some use scenarios, the scheduler is able to obtain extra knowledge about both the system (e.g. the real-time global UE AoI) and the channels (e.g. the channel SNR measurement). 
Leveraging such knowledge in scheduling will certainly further benefit the system in AoI reduction. More specifically, the original scheduling problem \eqref{prob:ofdm_scheduling}, with its MDP nature, can be transformed into a dynamic programming problem with the Bellman equation:

\begin{align}\label{eq:Bellman_equation}
		&\min\limits_{\{\mathbf{A}_\pi(t)\vert t\geqslant 1\}}\lim\limits_{T\to+\infty}\mathbb{E}\left(\frac{1}{IT}\sum\limits_{t=1}^T\sum\limits_{i\in\mathcal{I}}h_i(t)\right)\nonumber\\
		=&\lim\limits_{T\to+\infty}\frac{\min\limits_{\mathbf{A}(1)}\mathbb{E}\left(\sum\limits_{i\in\mathcal{I}}h_i(1)+\min\limits_{\{\mathbf{A}_\pi(t)\vert t\geqslant 2\}}\mathbb{E}\left(\sum\limits_{t=2}^T\sum\limits_{i\in\mathcal{I}}h_i(t)\right)\right)}{IT}\nonumber\\
		=&\lim\limits_{T\to+\infty}\frac{1}{IT}\left[\min\limits_{\mathbf{A}(1)}\mathbb{E}\left(\sum\limits_{i\in\mathcal{I}}h_i(1)+\min\limits_{\mathbf{A}(2)}\mathbb{E}\left(\sum\limits_{i\in\mathcal{I}}h_i(2)\right.\right.\right.\nonumber\\
		&+\left.\left.\left.\min\limits_{\{\mathbf{A}_\pi(t)\vert t\geqslant 3\}}\mathbb{E}\left(\sum\limits_{t=2}^T\sum\limits_{i\in\mathcal{I}}h_i(t)\right)\right)\right)\right]=\dots
\end{align}

In the extreme case where a perfect real-time knowledge in global UE AoI and CSI is available, even the global optimum of \eqref{eq:Bellman_equation} can be solved.} Nevertheless, it shall be remarked that such a global optimum can only be achieved through a joint optimization of $l$, $\mathcal{I}_t$ and $\mathbf{A}_t$, which relies on recursive approaches such like Block Coordinated Descent (BCD) or Sine Cosine Algorithm (SCA). Unfortunately, in practical scenarios where channels are fading, the convergence of such algorithms can probably take longer than the channel coherence time, which makes them impractical. Furthermore, the knowledge degree can be variant in time and space, varying among minimal, imperfect, and perfect levels. Therefore, taking a system design perspective, in this section we propose a three-stage resource scheduling framework, where heuristic enhancements can be attached to the optimal blind scheduler for further AoI reduction upon extra knowledge.

\subsection{Framework Design}
\begin{figure*}[!hbtp]
	\centering
	\includegraphics[width=.6\linewidth]{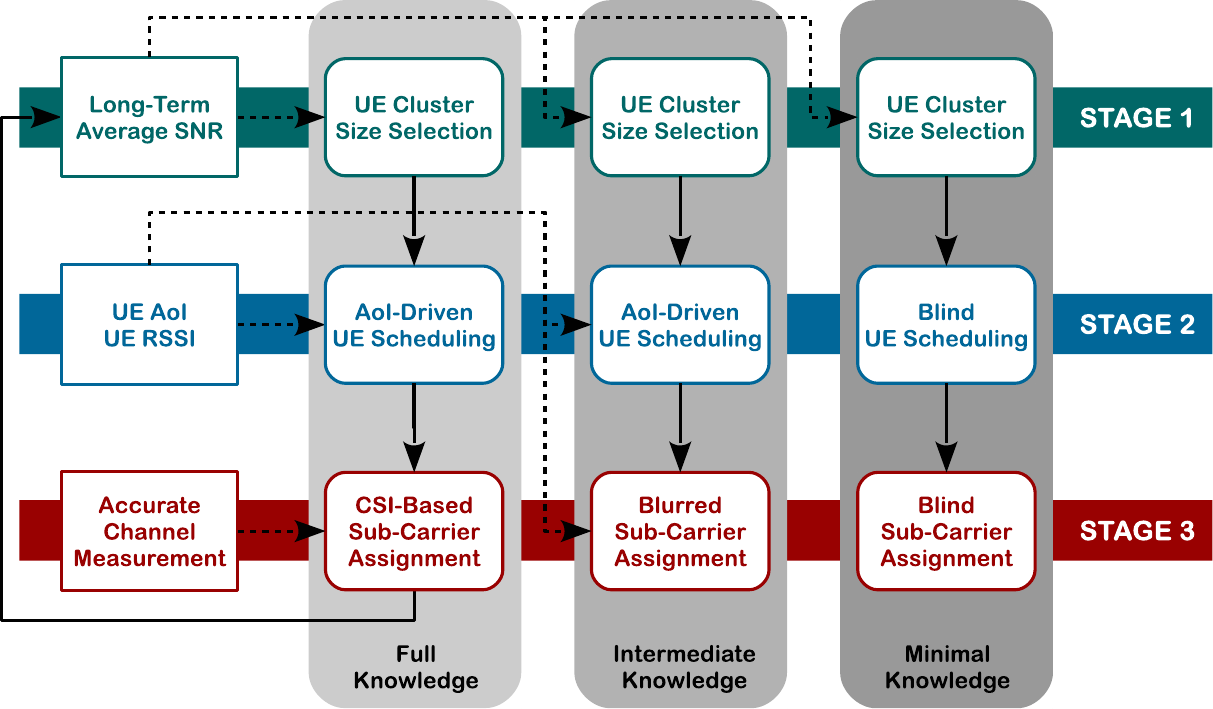}
	\caption{The proposed three-stage OFDMA resource scheduling framework.}
	\label{fig:three_stage_scheduling}
\end{figure*}

 Our proposed framework is illustrated in Fig.~\ref{fig:three_stage_scheduling}. The first stage focuses to select the cluster size $l$, requiring only the minimal knowledge about the long-term average PER among all UEs, as discussed in Sec.~\ref{subsec:cluster_size_selection}. According to the determined cluster size, in every slot $t$, on the second stage it selects $l$ UEs to commit message transmission, for which the instantaneous AoI and Received Signal Strength Indicator (RSSI) of every UE is available. Upon the selected UE cluster $\mathcal{I}_t$ in slot $t$, the scheduler assigns $M$ different sub-carriers to the $l$ UEs in a frequency-selective fashion, w.r.t. the accurate channel measurement of each UE. In case essential knowledge is unavailable, clustered RR and blind sub-carrier assignment can be deployed on stages 2 and 3, as we have discussed in Secs.~\ref{subsec:rr_optimum} and \ref{subsec:subcarrier_assignment}, respectively. With the system continuously working, the average PER will be online updated to enable long-term performance evolution.

\subsection{AoI-Driven Scheduling upon Transmission Feedback}
In use scenarios that allow the implementation of basic QoS management, an ACK/NACK feedback is usually available on the LLC layer for every transmission attempt, so that the server will be aware of the exact instantaneous AoI of every UE at arbitrary time. Furthermore, a rough estimation of the channel status, e.g. the RSSI, is also usually available for every individual UE in such scenarios. In this case, the intermediate level knowledge can be leveraged for a heuristic scheduling for AoI enhancement, which is based on AoI-index \cite{kadota18,jiang19_tcom}. \revise{More specifically,} in arbitrary time slot $t$, when assigned with a uniformly allocated blocklength $\sum\limits_{m\in\mathcal{M}}a_{\pi,m,i}\revise{(t)}=n\times \left\lfloor\frac{M}{l\subscript{opt}}\right\rfloor$, the PER $\varepsilon_{i}$ achievable by UE $i$ can be estimated by \eqref{eq:single_link_error_pro}. According to \cite{kadota18}, we compute the AoI index for every UE $i\in\mathcal{I}$, \revise{which is 
$\zeta_{i}(t)=h_{i}(t)\times\left[h_i(t)+1\right]\times(1-\varepsilon_{i})$}. In every time slot $t$, we propose to schedule the $l\subscript{opt}$ UEs with largest AoI indices to transmit.

\subsection{Recursive Sub-carrier Assignment upon Perfect Channel Status Information}\label{subsec:recursive_subcarrier_assignment}
\revise{If the channel is partially known, i.e., the accurate channel state of UE $i$ is given by
$\hat{h}_{m,i}=\rho h_{m,i}+\sqrt{1-\rho^2}\Delta h_{m,i}$, where $h_{m,i}$ is the imperfect channel knowledge, $\rho$ is the coherent factor and $\Delta h_{m,i}$ is an i.i.d. random variable. Then, denoting $f_H(\hat{h}_{m,i})$ the PDF of $\hat{h}_{m,i}$, the error probability in the FBL regime can be rewritten as
    \begin{align}
    \varepsilon_{i}{(t)}=&\int\left[ Q\left(\frac{\left(\sum\limits_{m\in\mathcal{M}}a_{\pi,m,i}{(t)}C_{m,i}(|\hat{h}_{m,i}|^2)n-\tau\right)\ln 2}{\sqrt{\sum\limits_{m\in\mathcal{M}}a_{\pi,m,i}{(t)}V_{m,i}(|\hat{h}_{m,i}|^2)n}}\right)\right.\nonumber\\
    &\left.\times\prod\limits_{m\in\mathcal{M}} f_H(\hat{h}_{m,i})\right]\diff\mathbf{Z},
    \end{align}    
with which the convexity of Problem~\eqref{problem_blocklength_blind} still hold. Therefore, it can still be solved efficiently. However, it is worth to mention that the sub-carrier assignment is no longer uniformly allocated. Moreover, to solve such convex problem requires a complexity of $\mathcal{O}(l^2)$.

On the other hand, if the full frequency-selective channel measurement} is available for all UEs in the current transmitting cluster, we can further improve the performance \revise{by exploiting a water-filling alike sub-carrier assignment.} Generally, we aim at an optimal assignment that balances the PERs among all UEs scheduled to transmit. \revise{This can be achieved by a recursive optimization algorithm with sorting and selection, which results a computational complexity of $\mathcal{O}(2Ml!\log l!)$. It is briefly summarized with the pseudo code in Fig.~\ref{alg:subcarrier_assignment}.}

\begin{figure}[!tb]
	\removelatexerror
	\begin{algorithm}[H]
		\footnotesize
		\DontPrintSemicolon
		Given $\mathcal{M}$, $l$, $\mathcal{I}_t$, and $\gamma_{m,i}$ for all $(m,i)\in\left(\mathcal{M}\times\mathcal{I}_t\right)$, where $\Vert\mathcal{I}_t\Vert_0=l$\;
		Initialization: randomly cluster $\mathcal{M}$ into $l$ subsets $\mathcal{M}_{i}$ where $i\in\mathcal{I}_t$\hspace{3.3cm}\hfill\emph{\raggedleft Random initial assignment}\;
		$\mathcal{M}'_i\gets\mathcal{M}_i,\quad\forall i\in\mathcal{I}_t$\;
		\While(\hfill\emph{Main iteration}){$\text{True}$}{
			Calculate $\varepsilon_{i}$ w.r.t. $\mathcal{M}_i$ according to \eqref{eq:single_link_error_pro}, for all $i\in\mathcal{I}_t$\;
			$i\subscript{best}\gets\arg\min\limits_{i\in\mathcal{I}_t}\varepsilon_{i}$, $i\subscript{worst}\gets\arg\max\limits_{i\in\mathcal{I}_t}\varepsilon_{i}$\hspace{4.3cm}\hfill\emph{\raggedleft Find UEs with lowest and highest PER}\;
			$i_\Delta\gets\arg\min\limits_{m\in\mathcal{M}_{i\subscript{best}}}\gamma_{i\subscript{best},m,t}$\;
			$\mathcal{M}'_{i\subscript{best}}=\mathcal{M}_{i\subscript{best}}-\{i_\Delta\}$, $\mathcal{M}'_{i\subscript{worst}}=\mathcal{M}_{i\subscript{worst}}+\{i_\Delta\}$\hspace{2.3cm}\hfill\emph{\raggedleft Re-assign the worst sub-carrier of $i\subscript{best}$ to $i\subscript{worst}$}\;
			Calculate $\varepsilon'_{i}$ w.r.t. $\mathcal{M}'_i$ according to \eqref{eq:single_link_error_pro}, for all $i\in\mathcal{I}_t$\;
			\uIf(\hfill\emph{Update the assignment if improved}){$\sum\limits_{i\in\mathcal{I}_t}\varepsilon'_{i}<\sum\limits_{i\in\mathcal{I}_t}\varepsilon_{i}$}{
				$\mathcal{M}_i\gets\mathcal{M}'_i,\quad\forall i\in\mathcal{I}_t$\;
			}
			\Else(\hfill\emph{Converged}){Break\;}
		}
		Return $\mathcal{M}_i$ for all $i\in\mathcal{I}_t$ as the sub-carrier assignment\;
	\end{algorithm}
	\caption{The recursive sub-carrier assignment algorithm upon perfect CSI}
	\label{alg:subcarrier_assignment}
\end{figure}

\section{Numerical Evaluation}\label{sec:simulations}

\subsection{Benchmarking Result: Cluster Size Determination}
To validate the cluster size selection we proposed in Sec.~\ref{subsec:cluster_size_selection}, we carried out a benchmarking test to compare its performance under with two baseline solutions. \revise{The first baseline aims at vanilla URLLC with a guaranteed PER up to $1\times10^{-5}$ within an air-interface delay up to \SI{10}{\milli\second},
this is achieved by solving
\begin{equation}
	l=\left\lfloor\frac{M}{\min\left( k\in\mathbb{N}^+ \left\vert Q\left(\frac{kC\subscript{avg}n\ln 2-\tau}{\sqrt{kV\subscript{avg}n}}\right)\leqslant 10^{-5}\right)\right.}\right\rfloor,
\end{equation}
where $C\subscript{avg}=\log_2(1+\bar{\gamma})$ and $V\subscript{avg}=1-\frac{1}{(1+\bar{\gamma})^2}$, and forcing $n\leqslant\left\lfloor\frac{\SI{10}{\milli\second}}{\SI{71.3}{\micro\second}}\right\rfloor=140$.
Here, \SI{71.3}{\micro\second} is the OFDM symbol length. If no positive root of $l$ exists even under $n=140$, the feasibility of URLLC mode is rejected -- which is not the case under any specification in our benchmark test.
The second is an un-clustered RR schedule that allocates all sub-carriers to only one UE in every slot, i.e. it simply takes $l=1$}. All three cluster-size selection strategies are combined with a round-robin scheduler and a random uniform sub-carrier assignment. 

\begin{figure*}[!htbp]
	\centering
	\begin{subfigure}[b]{.45\linewidth}
		\centering
		\includegraphics[width=\linewidth]{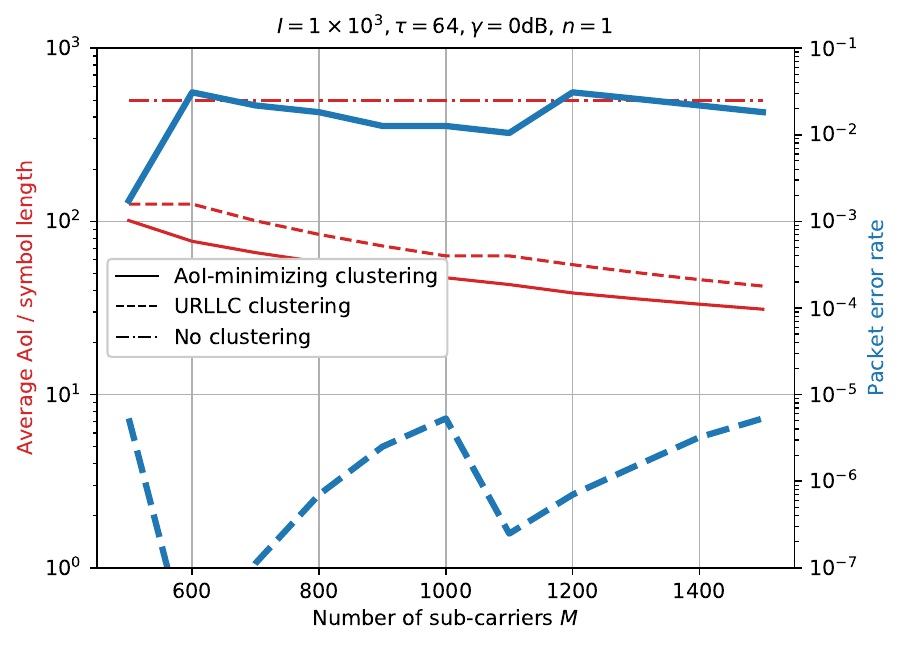}
	\end{subfigure}
	\begin{subfigure}[b]{.45\linewidth}
		\centering
		\includegraphics[width=\linewidth]{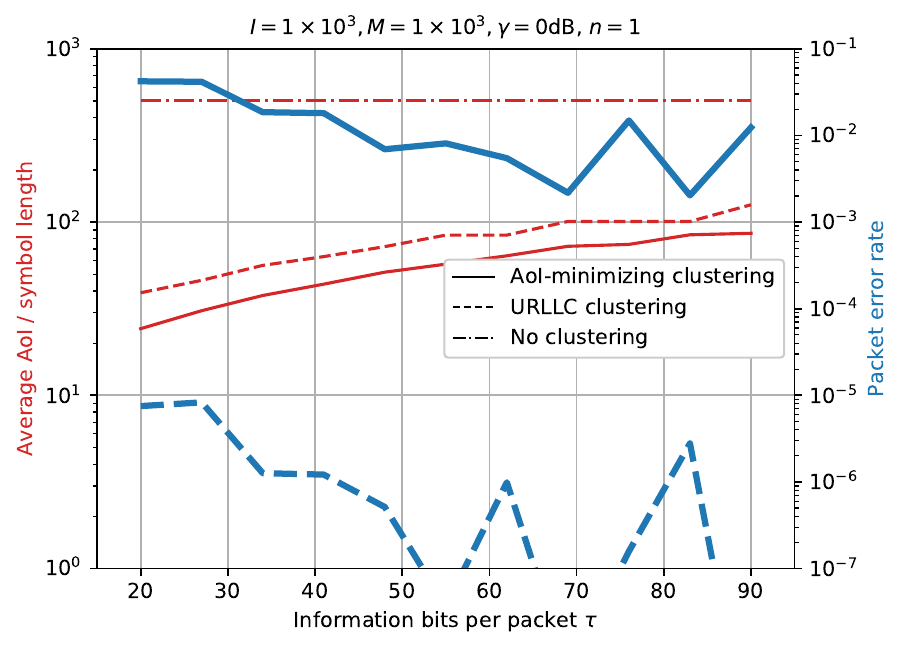}
	\end{subfigure}\\
	\begin{subfigure}[b]{.45\linewidth}
		\centering
		\includegraphics[width=\linewidth]{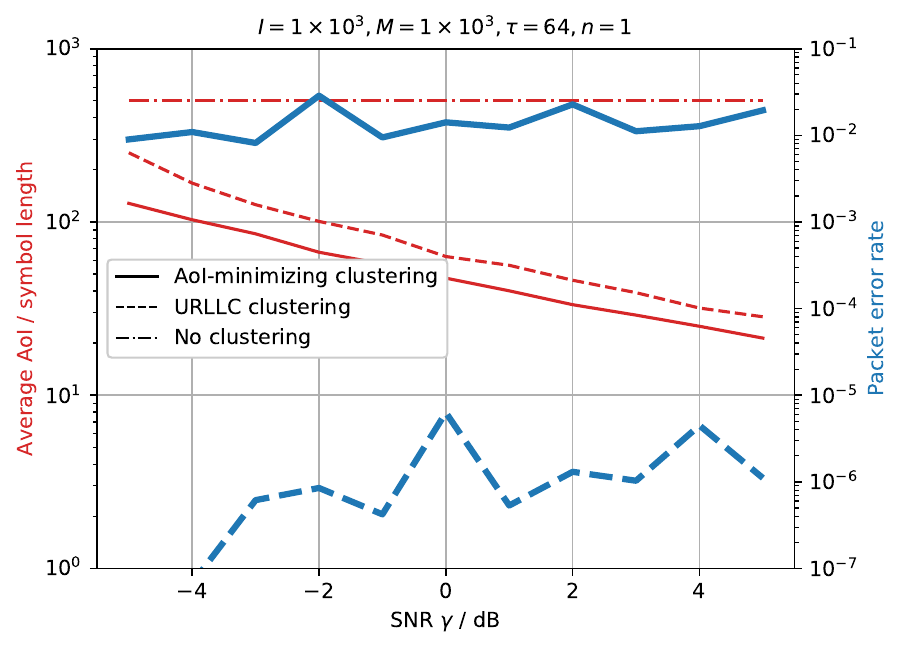}
	\end{subfigure}
	\begin{subfigure}[b]{.45\linewidth}
		\centering
		\includegraphics[width=\linewidth]{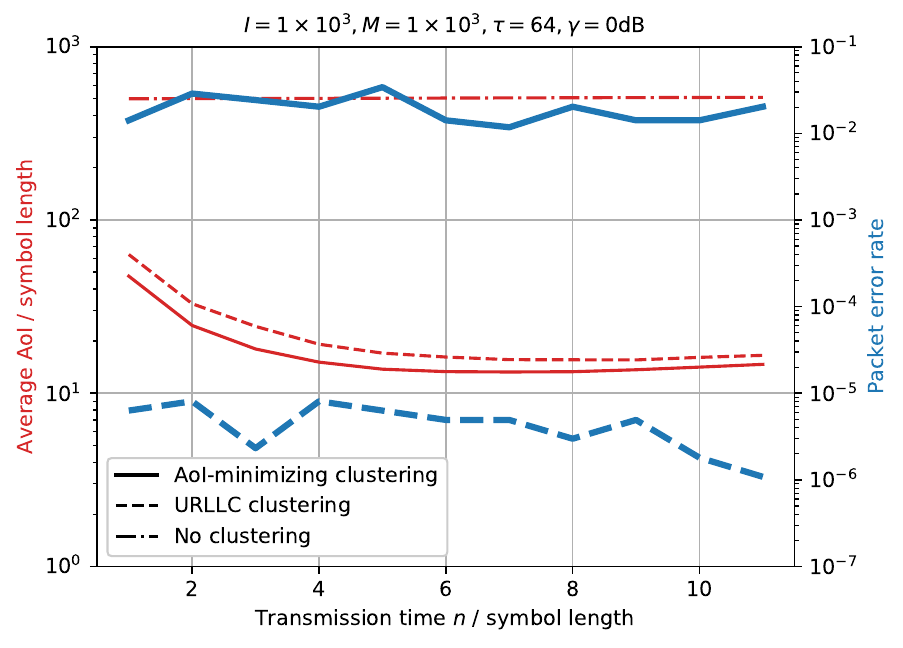}
	\end{subfigure}
	\caption{Benchmarking different clustering policies, regarding to both the average user AoI (red thin curves) and the PER (blue thick curves). Remark that the PER remains 0 under the no-clustering policy in all benchmark tests, and is therefore not appearing in the figures.}
	\label{fig:benchmark_clustering_policy}
\end{figure*}

We applied different specifications to investigate the sensitivities of referred solutions to: i) the number of sub-carriers $M$; ii) information bits per packet  $\tau$, iii) the average UE SNR $\bar{\gamma} $, and iv) the transmission slot length $n$. For each specification and each solution, $10~000$ Monte-Carlo tests were executed. The results are illustrated in Fig.~\ref{fig:benchmark_clustering_policy}, where the proposed AoI-minimizing clustering strategy significantly outperforms both the baselines in AoI reduction under all specifications - even though it suffers from a much higher PER than that of the vanilla URLLC strategy.


\subsection{Gains of Heuristic Enhancers}
\begin{figure*}[!hbtp]
	\centering
	\includegraphics[width=.7\linewidth]{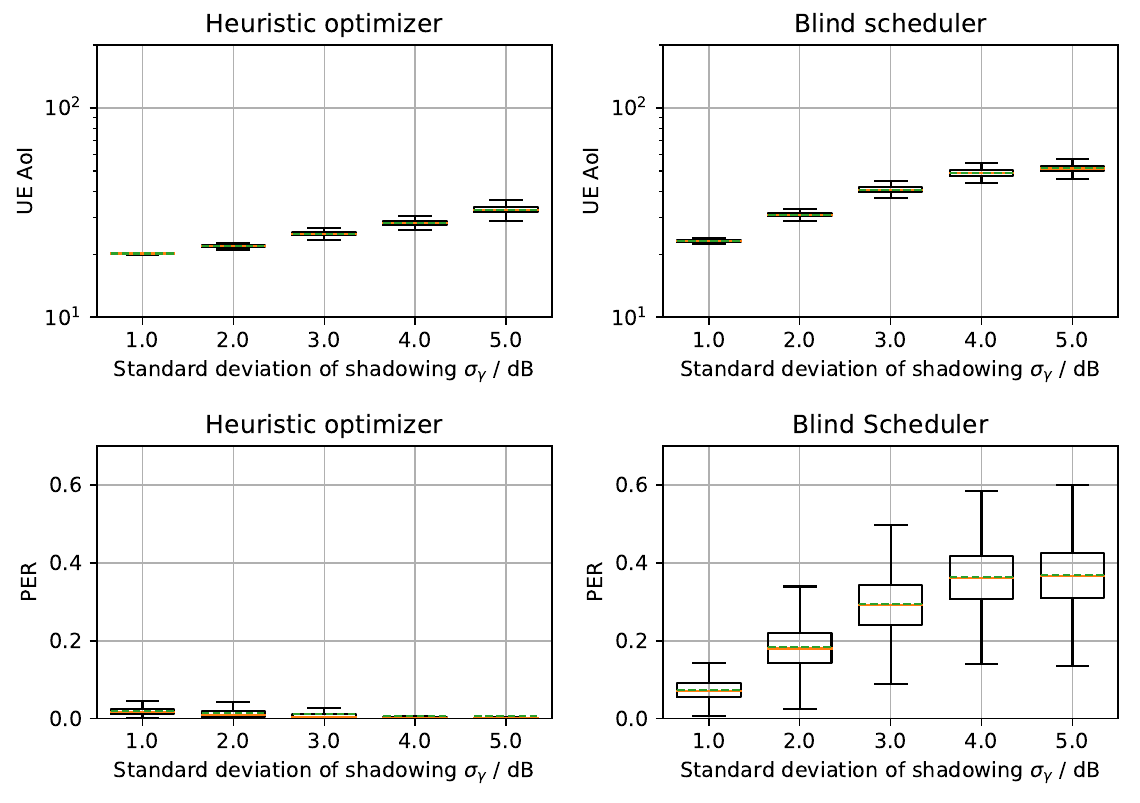}
	\caption{The heterogeneity of UEs and Rayleigh fading bring channel diversities -- in domains of space, frequency, and time -- to the system, which can be exploited by our recursive sub-carrier assignment mechanism to generate diversity gain.}
	\label{fig:sim_fading_ch}
\end{figure*}

To evaluate the performance gain that can be created by the heuristic enhancers leveraging extra knowledge of AoI and CSI, we conducted system-level simulations under Rayleigh faded, log-normal shadowed heterogeneous Gaussian channels for different UEs. The system and environment are specified as listed in Tab.~\ref{tab:sim_spec}.
\begin{table}[!hbtp]
	\centering
	\begin{tabular}{l|l}
		\toprule[2px]
		UE number $I$ & $1000$\\
		Sub-carrier number $M$ & $1000$\\
		Slot length $\tau$ (in symbol length) & $1$\\
		Reference SNR $\bar\gamma$ (without fading) & \SI{5}{\dB}\\
		Log-normal shadowing variance $\sigma_\gamma$& $1\sim 5$\si{\dB}\\
		Average Rayleigh fading duration& \SI{5}{\milli\second}\\
		Coherence bandwidth of Rayleigh fading&\SI{900}{\kilo\hertz}\\
		Rayleigh fading scale & 1\\
		Sub-carrier bandwidth & \SI{15}{\kilo\hertz}\\
		OFDM symbol length (with cyclic prefix) & \SI{71.3}{\micro\second}\\		
		\bottomrule[2px]
	\end{tabular}
	\caption{Specifications of the system-level simulation}
	\label{tab:sim_spec}
\end{table}
Under each specification of shadowing effect, we repeated $10~000$ times Monte-Carlo test. In each test we first evaluated the proposed solution in the minimal knowledge case, where a blind clustered RR scheduler was applied with the AoI-optimal cluster size and combined with a random uniform sub-carrier assignment. Then we granted the system with full knowledge about instantaneous UE AoI and perfect CSI, so that the AoI index based scheduling and recursive sub-carrier assignment mechanisms are activated. The overall results are plotted in Fig.~\ref{fig:sim_fading_ch}. Generally we can observe a significant gain generated by the heuristic enhancements, in both aspects of the AoI and PER. It is worth to remark that the AoI performance of blind scheduler is satisfactory even under strong shadowing effect, despite of its high PER sensitivity to the shadowing variance (which is due to the poor performance of blind sub-carrier allocator under heterogeneous and frequency selective channel conditions).

\section{Conclusion}\label{sec:conclusion}
In this paper, we have studied the problem of blind OFDMA scheduling towards average AoI minimization in the FBL regime. When the system knowledge is minimized, i.e. only the long-term average PER among all UEs and the individual transmission schedule history are available, our analyses have revealed the optimal blind resource scheduling policy, which is assembled by a clustered round-robin scheduler in then time domain, and a random uniform sub-carrier assignment in the frequency domain. We have also analytically solved the optimal cluster size 
given the average PER, which is independent from the total UE amount. As an enhancing solution to extend our proposed approach to more general scenarios where additional knowledge about system and channel may be available, we have designed a three-stage scheduling framework to support the deployment of heuristic optimization of AoI. With numerical simulations, we have demonstrated the superiority of our proposed optimal clustering policy over benchmark solutions, and the effectiveness of heuristic enhancements.

\appendices

\section{\revise{Proofs of Theorem~\ref{th:blind_optimum_rr} and Corollary~\ref{cr:blind_optimum_crr}}}\label{app:opt_rr_proof}
\begin{proof}[Proof of Theorem~\ref{th:blind_optimum_rr}]
	First, define the system state of a blind schedule $\pi$ as a vector of vectors $\mathbf{S}(\pi,t)=\left(\mathbf{s}_1^{(\pi,t)},\mathbf{s}_2^{(\pi,t)},\dots,\mathbf{s}_I^{(\pi,t)}\right)\transpose$ where each element vector $\mathbf{s}_i^{(\pi,t)}$ contains the full history of (ITI) of UE $i$ till transmission slot $t$. Without loss of generality we consider the length \revise{of} every  transmission slot (in symbol length) $n=1$, and first investigate the case of zero initial state
	\begin{equation}
	\mathbf{s}_i^{(\pi,1)}=(1),\quad\forall i\in\mathcal{I}.
	\end{equation}
	For simplification of notation we let $k(\pi,i,t)=\left\Vert\mathbf{s}_i^{(\pi,t)}\right\Vert_0$, then after every transmission slot, $\mathbf{S}$ is updated row-by-row w.r.t. the schedule $\mathcal{I}_t$:
	\begin{align}
		&\mathbf{s}_i^{(\pi,t+1)}\\
		=&\begin{cases}
			\left(s_{i,1}^{(\pi,t)},s_{i,2}^{(\pi,t)},\dots,s_{i,k(\pi,i,t)}^{(\pi,t)},1\right)& i\in\mathcal{I}_t,\\
			\left(s_{i,1}^{(\pi,t)},s_{i,2}^{(\pi,t)},\dots,s_{i,k(\pi,i,t)-1}^{(\pi,t)},s_{i,k(\pi,i,t)}^{(\pi,t)}+1\right)& \text{otherwise}.
		\end{cases}\nonumber
	\end{align}
	\revise{For all $\forall(i,t)\in\left(\mathcal{I}\times\mathbb{N}^+\right)$,} $\sum\limits_{\kappa=1}^{k(\pi,i,t)}s_{i,\kappa}^{(\pi,t)}=t$ and $\sum\limits_{i\in\mathcal{I}}k(\pi,i,t)=I+l(t-1)$. Generalizing \revise{it} into
	\begin{equation}
	\tilde{s}_{i,j}^{(\pi,t)}=\begin{cases}
	s_{i,j}^{(\pi,t)}&1\le j\le k(\pi,i,t),\\
	0&\text{otherwise},
	\end{cases}
	\end{equation}
	we can obtain that the AoI expectation of UE $i$ after $t$ slots under schedule $\pi$ is
	\begin{equation}
	\begin{split}
		&\bar h_{i,\pi}^{(t,I)}=\sum\limits_{\kappa=1}^{k(\pi,i,t)}s_{i,\kappa}^{(\pi,t)}\theta_{i,k(\pi,i,t)-\kappa}^{(\pi,t)}\\
		=&\sum\limits_{\kappa=0}^{k(\pi,i,t)}\tilde{s}_{i,\kappa}^{(\pi,t)}\theta_{i,k(\pi,i,t)-\kappa}^{(\pi,t)}=\sum\limits_{\kappa=0}^{k(\pi,i,t)}\tilde{s}_{i,k(\pi,i,t)-\kappa}^{(\pi,t)}\theta_{i,\kappa}^{(\pi,t)},
	\end{split}
	\end{equation}
	where
	\begin{equation}
		\theta_{i,\kappa}^{(\pi,t)}\triangleq\begin{cases}
			1&\kappa=0\\
			\prod\limits_{p=1}^\kappa\varepsilon_i\left(\sum\limits_{q=1}^p\tilde{s}_{i,q}^{(\pi,t)}\right)&\kappa\in\mathbb{N}^+
		\end{cases}.
	\end{equation}
	And the expected average AoI among all UEs is
	\begin{equation}
		\begin{split}
			&\bar h_{\pi}^{(t,I)}=\mathbb{E}\left\{\frac{1}{I}\sum\limits_{i\in\mathcal{I}}{\bar h_{i,\pi}^{(t,I)}}\right\}\\
			=&\frac{1}{I}\left\{\sum\limits_{i\in\mathcal{I}}\sum\limits_{\kappa=0}^{k(\pi,i,t)}\tilde{s}_{i,k(\pi,i,t)-\kappa}^{(\pi,t)}\theta_{i,\kappa}^{(\pi,t)}\right\}.
		\end{split}
	\end{equation}
	Since $\varepsilon_{i}(t)$ is i.i.d. for all $(i,t)$ with the expectation $\mathbb{E}\{\varepsilon\}=\lim\limits_{T\to+\infty}\frac{1}{TI}\sum\limits_{t=1}^T\sum\limits_{i\in\mathcal{I}}\varepsilon_{i}(t)=\varepsilon\subscript{avg}$, we have $\mathbb{E}\left\{\theta_{i,\kappa}^{(\pi,t)}\right\}=\prod\limits_{p=1}^\kappa\varepsilon\subscript{avg}=\varepsilon\subscript{avg}^\kappa$ for all $\kappa\in\mathbb{N}^+$, and therefore
	\begin{equation}
		\begin{split}
		&\bar h_{\pi}^{(t,I)}=\frac{1}{I}\sum\limits_{i\in\mathcal{I}}\sum\limits_{\kappa=0}^{k(\pi,i,t)}\tilde{s}_{i,k(\pi,i,t)-\kappa}^{(\pi,t)}\mathbb{E}\left\{\theta_{i,\kappa}^{(\pi,t)}\right\}\\
		=&\frac{1}{I}\sum\limits_{i\in\mathcal{I}}\sum\limits_{\kappa=0}^{k(\pi,i,t)}\tilde{s}_{i,k(\pi,i,t)-\kappa}^{(\pi,t)}\varepsilon\subscript{avg}^\kappa.\label{eq:mean_aoi_under_schedule}
		\end{split}
	\end{equation}	

	Now we define $K(\pi,t)=\max\limits_{i\in\mathcal{I}}k(\pi,i,t)$,
	and construct
	\begin{equation}
	\mathfrak{s}^{(\pi,t)}_\kappa=\frac{1}{tI}\sum\limits_{i\in\mathcal{I}}\tilde{s}_{i,k(\pi,i,t)-\kappa}^{(\pi,t)},
	\end{equation}
	so that \eqref{eq:mean_aoi_under_schedule} becomes $\bar h_{\pi}^{(t,I)}=t\sum\limits_{\kappa=0}^{K(\pi,t)}\mathfrak{s}_\kappa^{(\pi,t)}\varepsilon\subscript{avg}^\kappa$.
	
	For all $(\pi,t)$ we construct a tuple $\left(\mathcal{K}_t,\mathcal{F}_t,P_{\pi,t}\right)$ where $\mathcal{K}_t\triangleq[1,2,\dots,t]$, $\mathcal{F}_t$ is the power set of $\mathcal{K}_t$,
	and $P_{\pi,t}$ is a measure on $\mathcal{F}_t$. Here we use $\mathfrak{s}^{(\pi,t)}_\kappa$ to construct such a $P_{\pi,t}$ that
	$P_{\pi,t}(\kappa)=\mathfrak{s}^{(\pi,t)}_\kappa$.
	Note that such a tuple matches the definition of probability space, since $\mathfrak{s}^{(\pi,t)}_\kappa\geqslant 0$ for all $\kappa\in\mathbb{N}$, and $\sum\limits_{\kappa\in\mathbb{N}}\mathfrak{s}^{(\pi,t)}_\kappa=\sum\limits_{\kappa\in\mathcal{K}(\pi,t)}\mathfrak{s}^{(\pi,t)}_\kappa=1$.
	Thus, we can define the corresponding discrete cumulative distribution function (CDF):
	\begin{equation}
	\mathfrak{S}^{(\pi,t)}_\kappa\triangleq\sum\limits_{j=0}^{\kappa}\mathfrak{s}_j^{(\pi,t)}
	\end{equation}
	For two different probabilistic measurements with cumulative distribution functions $F_1(x)$ and $F_2(x)$, respectively, over the same sample space $x\in\mathcal{X}$, $F_1$ second-order stochastically dominates $F_2$, if and only if $\int\limits_{-\infty}^xF_1(y)\diff y\le \int\limits_{-\infty}^xF_2(y)\diff y,\forall x\in\mathcal{X}$ \cite{levy2015stochastic}. Therewith, for two different transmission schedulers $\pi$ and $\pi'$, we call $\pi$ to second-order statistically dominate $\pi'$ at $t$, if and only if \revise{$\sum\limits_{k=0}^\kappa\mathfrak{S}^{(\pi,t)}_k\le\sum\limits_{k=0}^\kappa\mathfrak{S}^{(\pi',t)}_k$ for all $\kappa\in\mathcal{K}_t$}.
	
	Now we investigate the second-order stochastic dominance between a pair of different schedules $(\pi,\pi')$ that differ from each other only by one transmission slot. More specifically, in slot $\tau$, $\pi$ and $\pi'$ assigns UEs $x$ and $y$ to transmit, respectively, while all the rest slots $t\neq\tau$ are assigned in the same way for both the schedules. We denote the transmission attempts made by UEs $x$ and $y$ till $\tau$ under schedule $\pi$ as $k_x$ and $k_y$, respectively, thus:
	\begin{align}
	&\sum\limits_{k=1}^{k_y-1}\tilde{s}_{y,k}^{(\pi,t)}<\underset{=\tau}{\underbrace{\sum\limits_{k=1}^{k_x}\tilde{s}_{x,k}^{(\pi,t)}}}<\sum\limits_{k=1}^{k_y}\tilde{s}_{y,k}^{(\pi,t)},\label{eq:slot_assignment_shift_constraint_1}\\
	&\tilde{s}_{i,k}^{(\pi,t)}=\tilde{s}_{h,k}^{(\pi',t)},\quad\forall{i\in\left(\mathcal{I}-\{x,y\}\right)},\forall k\\
	&\tilde{s}_{x,k}^{(\pi',t)}=\begin{cases}
		\tilde{s}_{x,k}^{(\pi,t)}&\revise{k}<k_x\\
		\tilde{s}_{x,k_x}^{(\pi,t)}+\tilde{s}_{x,k_x+1}^{(\pi,t)}&k=k_x\\
		\tilde{s}_{x,\revise{k}+1}^{(\pi,t)}&\revise{k}>k_x
	\end{cases}\\
	&\tilde{s}_{y,k}^{(\pi',t)}=\begin{cases}
	\tilde{s}_{y,k}^{(\pi,t)}&k<k_y\\
	\tilde{s}_{y,k-1}^{(\pi,t)}&k>k_y+1
	\end{cases}\\
	&\tilde{s}_{y,k_y}^{(\pi',t)}+\tilde{s}_{y,k_y+1}^{(\pi',t)}=\tilde{s}_{y,k_y}^{(\pi,t)}\\
	&k(\pi',x,t)=k(\pi,x,t)-1\\
	&k(\pi',y,t)=k(\pi,y,t)+1
	\end{align}
	Now given any certain time $t$, we compare the second-order CDFs $\sum\limits_{k=0}^\kappa\mathfrak{S}^{(\pi,t)}_k$ and $\sum\limits_{k=0}^\kappa\mathfrak{S}^{(\pi',t)}_k$:
	
	\begin{equation}
	\begin{split}
	&\sum\limits_{k=0}^\kappa\mathfrak{S}^{(\pi,t)}_k-\sum\limits_{k=0}^\kappa\mathfrak{S}^{(\pi',t)}_k\\
	=&\frac{1}{tI}\sum\limits_{k=0}^\kappa\sum\limits_{j=1}^{k}\sum\limits_{i\in\mathcal{I}}\left[\tilde{s}_{i,k(\pi',i,t)-j}^{(\pi',t)}-\tilde{s}_{i,k(\pi,i,t)-j}^{(\pi,t)}\right]\\
	=&\frac{1}{tI}\sum\limits_{k=0}^\kappa\sum\limits_{j=1}^k\left[\tilde{s}_{x,k(\pi',x,t)-j}^{(\pi',t)}+\tilde{s}_{y,k(\pi',y,t)-j}^{(\pi',t)}\right.\\
	&\left.-\tilde{s}_{x,k(\pi,x,t)-j}^{(\pi,t)}-\tilde{s}_{y,k(\pi,y,t)-j}^{(\pi,t)}\right]
	\end{split}
	\end{equation}
	
	Next, given any $(t,\pi,\kappa,x,y)$, we design the following operation of transmission slot exchange:
	\begin{enumerate}
		\item Find the time slot $\tau_1$ that $\sum\limits_{j=1}^{k(\pi,x,t)-\kappa}\tilde{s}_{x,j}^{(\pi,t)}=\tau_1$, i.e. $x$ is scheduled to make its $\left(k(\pi,x,t)-\kappa\right)^\text{th}$ transmission attempt at $\tau_1$.
		\item Since $x$ is transmitted at $\tau_1$, $y$ must be idle at $\tau_1$, hence $\exists\kappa'$ that $\sum\limits_{j=1}^{k(\pi,y,t)-\kappa'-1}\tilde{s}_{y,j}^{(\pi,t)}<\tau_1<\sum\limits_{j=1}^{k(\pi,y,t)-\kappa'}\tilde{s}_{y,j}^{(\pi,t)}$, update $\pi$ to $\pi''$ by reassigning the slot $\tau_1$ from $x$ to $y$ for transmission. 
		\item If $k(\pi,y,t)\geqslant\kappa+1$, let $\tau_2=\sum\limits_{j=1}^{k(\pi,y,t)-\kappa}\tilde{s}_{y,j}^{(\pi,t)}$, update $\pi''$ to $\pi'$ by reassigning the slot $\tau_2$ from $y$ to $x$ for transmission. Otherwise, $\tau_2=0$ and $\pi'=\pi''$
		\item Return $\pi'$.
	\end{enumerate}
	
	Since $\pi$ differs from $\pi''$ only by one transmission slot at $\tau_1$, while $\pi''$ only differs from $\pi'$ by another transmission slot at $\tau_2$, we can obtain that\cleardoublepage
	
	\begin{strip}
	\vspace{-3mm}
	\begin{equation}
	\begin{split}
	&\sum\limits_{k=0}^\kappa\mathfrak{S}^{(\pi,t)}_k-\sum\limits_{k=0}^\kappa\mathfrak{S}^{(\pi',t)}_k=\sum\limits_{k=0}^\kappa\mathfrak{S}^{(\pi,t)}_k-\sum\limits_{k=0}^\kappa\mathfrak{S}^{(\pi'',t)}_k+\sum\limits_{k=0}^\kappa\mathfrak{S}^{(\pi'',t)}_k-\sum\limits_{k=0}^\kappa\mathfrak{S}^{(\pi',t)}_k\\
	=&\frac{1}{tI}\sum\limits_{k=1}^\kappa\sum\limits_{j=1}^k\left[\tilde{s}_{x,k(\pi',x,t)-j}^{(\pi',t)}+\tilde{s}_{y,k(\pi',y,t)-j}^{(\pi',t)}-\tilde{s}_{x,k(\pi,x,t)-j}^{(\pi,t)}-\tilde{s}_{y,k(\pi,y,t)-j}^{(\pi,t)}\right]
	\end{split}	
	\end{equation}

	
	For convenience of notation we define that $k_x\triangleq k(\pi,x,t)$, $k_y\triangleq k(\pi,y,t)$, $k_x'\triangleq k(\pi',x,t)$, $k_y'\triangleq k(\pi',y,t)$, and $\kappa''\triangleq\arg_\revise{J}\left\{\sum_{j=\revise{J}}^{k(\pi',x,t)}\tilde{s}^{\pi',t}_{x,j}=t-\tau_2\right\}$. 
	First we force $\kappa'\leqslant\kappa\revise{''}-2$, so that this exchange guarantees, as illustrated in Fig.~\ref{fig:slot_exchange}:
	\vspace{1cm}
	\begin{equation}
	\mathfrak{s}^{(\pi',t)}_k=\begin{cases}
	\mathfrak{s}^{(\pi,t)}_k&0\le k\le\kappa'-1\\
	\mathfrak{s}^{(\pi,t)}_{k}-\frac{1}{tI}\tilde{s}^{\pi',t}_{k_y'-\kappa'-1}&k=\kappa'\\
	\mathfrak{s}^{(\pi,t)}_{k}+\frac{1}{tI}\tilde{s}^{\pi',t}_{y,k_y'-\kappa'-1}-\frac{1}{tI}\tilde{s}^{(\pi,t)}_{y,k_y-\kappa'-1}&k=\kappa'+1\\
	\mathfrak{s}^{(\pi,t)}_{k}+\frac{1}{tI}\tilde{s}^{(\pi,t)}_{y,k_y-k}-\frac{1}{tI}\tilde{s}^{(\pi,t)}_{y,k_y-k-1}&\kappa'+1<k<\kappa-1\\
	\mathfrak{s}^{(\pi,t)}_{k}+\frac{1}{tI}\tilde{s}^{(\pi,t)}_{y,k_y-\kappa+1}-\frac{1}{tI}\tilde{s}^{(\pi,t)}_{y,k_y-\kappa}+\frac{1}{tI}\tilde{s}^{(\pi,t)}_{x,k_x-\kappa}&k=\kappa-1\\
	\mathfrak{s}^{(\pi,t)}_{k}+\frac{1}{tI}\tilde{s}^{(\pi,t)}_{y,k_y-\kappa}-\frac{1}{tI}\tilde{s}^{(\pi,t)}_{x,k_x-\kappa}+\frac{1}{tI}\tilde{s}^{(\pi,t)}_{x,k_x-\kappa-1}&k=\kappa\\
	\mathfrak{s}^{(\pi,t)}_{k}+\frac{1}{tI}\tilde{s}^{(\pi,t)}_{x,k_x-k-1}-\frac{1}{tI}\tilde{s}^{(\pi,t)}_{x,k_x-k}&\kappa<k<\kappa''-1\\
	\mathfrak{s}^{(\pi,t)}_{k}+\frac{1}{tI}\tilde{s}^{(\pi',t)}_{x,k_x-\kappa''+1}-\frac{1}{tI}\tilde{s}^{(\pi,t)}_{x,k_x-\kappa''+1}&k=\kappa''-1\\
	\mathfrak{s}^{(\pi,t)}_{k}-\frac{1}{tI}\tilde{s}^{(\pi',t)}_{x,k_x-\kappa''+1}&k=\kappa''\\
	\mathfrak{s}^{(\pi,t)}_k&k>\kappa''
	\end{cases},
	\end{equation}
	which leads to
	\begin{equation}
	\mathfrak{S}^{(\pi',t)}_k=\begin{cases}
	\mathfrak{S}^{(\pi,t)}_k&0\le k\le\kappa'-1\\
	\mathfrak{S}^{(\pi,t)}_{k}-\frac{1}{tI}\tilde{s}^{(\pi,t)}_{y,k_y-k-1}&\kappa'\le k<\kappa-1\\
	\mathfrak{S}^{(\pi,t)}_{k}-\frac{1}{tI}\tilde{s}^{(\pi,t)}_{y,k_y-\kappa}+\frac{1}{tI}\tilde{s}^{(\pi,t)}_{x,k_x-\kappa}&k=\kappa-1\\
	\mathfrak{S}^{(\pi,t)}_{k}+\frac{1}{tI}\tilde{s}^{(\pi,t)}_{x,k_x-k-1}&\kappa\le k<\kappa''-1\\
	\mathfrak{S}^{(\pi,t)}_{k}+\frac{1}{tI}\tilde{s}^{(\pi',t)}_{x,k_x-\kappa''+1}&k=\kappa''-1\\
	\mathfrak{S}^{(\pi,t)}_{k}&k\geqslant\kappa''
	\end{cases},
	\end{equation}
	and therefore
	\begin{equation}
	-tI\left[\sum\limits_{k=0}^\kappa\mathfrak{S}^{(\pi',t)}_k-\sum\limits_{k=0}^\kappa\mathfrak{S}^{(\pi,t)}_k\right]
	=\begin{cases}
	0&0\le k\le\kappa'-1\\
	\tilde{s}^{(\pi',t)}_{k_y'-\kappa'-1}&k=\kappa'\\
	\tilde{s}^{(\pi',t)}_{k_y'-\kappa'-1}+\sum\limits_{j=\kappa'}^{k}\tilde{s}^{(\pi,t)}_{y,k_y-j-1}&\kappa'+1\le k<\kappa-1\\
	\tilde{s}^{(\pi',t)}_{k_y'-\kappa'-1}+\sum\limits_{j=\kappa'}^{\kappa}\tilde{s}^{(\pi,t)}_{y,k_y-j-1}-\sum\limits_{j=\kappa-1}^{k}\tilde{s}^{(\pi,t)}_{x,k_x-j-1}&\kappa-1\le k<\kappa''-1\\
	\tilde{s}^{(\pi',t)}_{k_y'-\kappa'-1}+\sum\limits_{j=\kappa'}^{\kappa}\tilde{s}^{(\pi,t)}_{y,k_y-j-1}-\sum\limits_{j=\kappa-1}^{\kappa''-2}\tilde{s}^{(\pi,t)}_{x,k_x-j-1}-\tilde{s}^{(\pi',t)}_{x,k_x-\kappa''+1}&k\geqslant \kappa''-1
	\end{cases}.
	\end{equation}
	Noticing 
	\begin{equation}
		\begin{split}
			&\tilde{s}^{(\pi',t)}_{k_y'-\kappa'-1}+\sum\limits_{j=\kappa'}^{\kappa}\tilde{s}^{(\pi,t)}_{y,k_y-j-1}
			=\sum\limits_{j=\kappa-1}^{\kappa''-2}\tilde{s}^{(\pi,t)}_{x,k_x-k-1}+\tilde{s}^{(\pi',t)}_{x,k_x-\kappa''+1}
			=\tau_2-\tau_1>0,
		\end{split}
	\end{equation} 
	\revise{we always have}
	\begin{align}
		&\sum\limits_{k=0}^\kappa\mathfrak{S}^{(\pi',t)}_k-\sum\limits_{k=0}^\kappa\mathfrak{S}^{(\pi,t)}_k\le0,\quad \kappa\in\mathbb{N},\label{eq:single_slot_assignment_swap_dominance}\\
		&\sum\limits_{k=0}^{+\infty}\mathfrak{s}^{(\pi',t)}_kk-\sum\limits_{k=0}^{+\infty}\mathfrak{s}^{(\pi,t)}_kk		=\tilde{s}_{y,k_y-\kappa+1}^{\pi,t}+\sum_{k=k'_y-\kappa+1}^{k'_y-\kappa'-1}\tilde{s}_{y,k}^{\pi',t}-\tilde{s}_{x,k'_x-\kappa''}^{\pi',t}+\sum_{k=k_x-\kappa''}^{k_x-\kappa}\tilde{s}_{x,k}^{\pi,t}
		=\left(\tau_1-\tau_2\right)-\left(\tau_1-\tau_2\right)=0,
	\end{align}
	\end{strip}
	\noindent which \revise{imply} that $\pi'$ second-order stochastically dominates $\pi$. A simple numerical example can be constructed as follows: let $\mathcal{I}=\{1,2\}$, and the transmission schedule $\pi$ be listed in Tab.~\ref{tab:num_example_slot_exchange}, for $t=10$ we can find the tuple $x=1, y=2, \kappa=3, \kappa'=2, \kappa''=4$ to execute the slot exchange, which generates a $\pi'$ that is also listed in Tab.~\ref{tab:num_example_slot_exchange}. The inter-transmission intervals and the stochastic measures in $\kappa$ domain of both schedules $\pi$ and $\pi'$ are illustrated in Fig.~\ref{fig:num_example_slot_exchange}, implying a second-order stochastic dominance preferring $\pi'$ over $\pi$.
	\begin{figure*}[!hbtp]
		\centering
		\includegraphics[width=.7\linewidth]{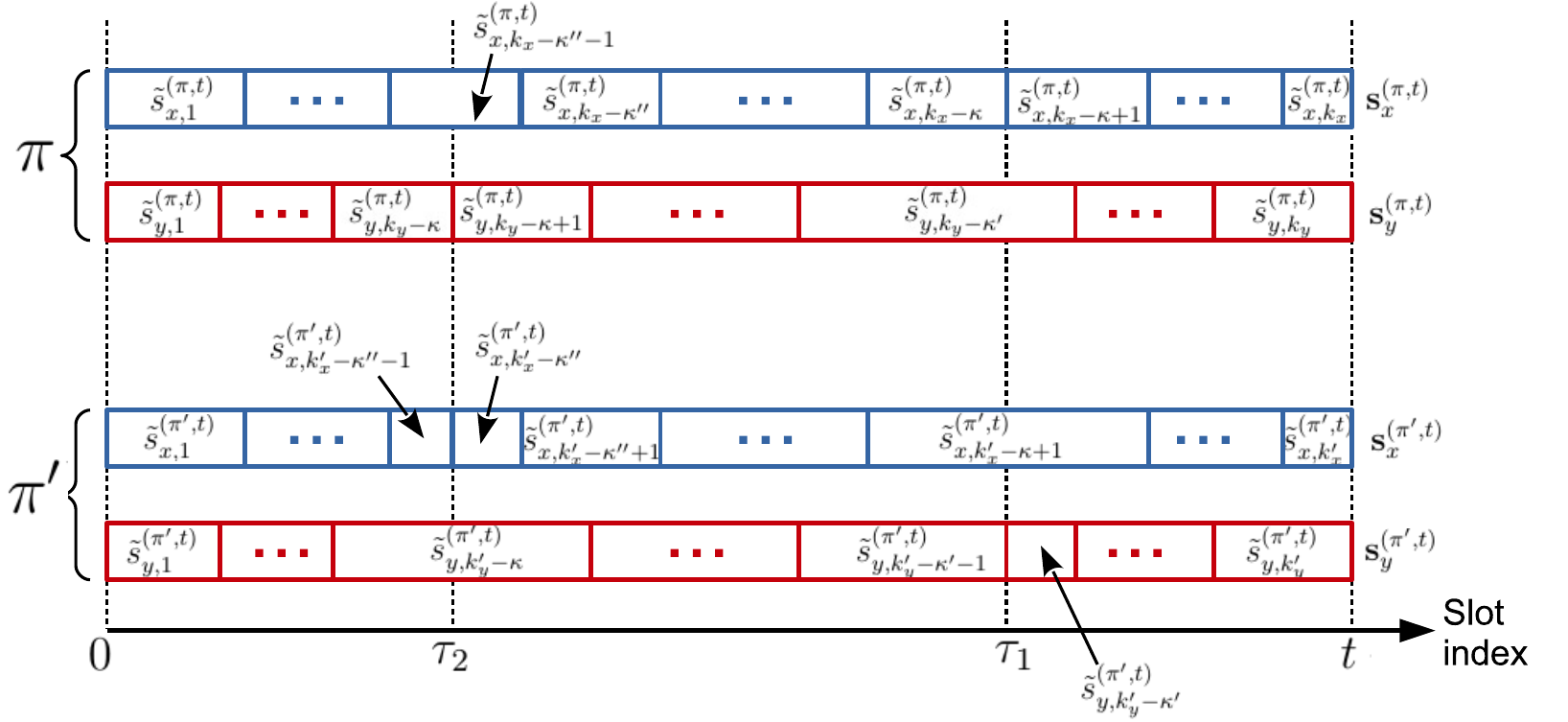}
		\caption{Update of involved ITIs through the proposed transmission slot exchange operation.}
		\label{fig:slot_exchange}
	\end{figure*}
	\begin{table}[!htbp]
		\centering
		\caption{A simple numerical example of transmission slot exchange}
		\label{tab:num_example_slot_exchange}
		\begin{tabular}{l|cccccccccc}
			\toprule[2px]
			Transmission slot&1&2&3&4&5&6&7&8&9&10\\
			\midrule[1px]
			Scheduled UE ($\pi$)&1&2&2&\textbf{2}&1&\textbf{1}&2&1&2&1\\
			Scheduled UE ($\pi'$)&1&2&2&\textbf{1}&1&\textbf{2}&2&1&2&1\\
			\bottomrule[2px]
		\end{tabular}
	\end{table}
	\begin{figure}[!htbp]
		\centering
		\begin{subfigure}[b]{\linewidth}
			\centering
			\includegraphics[width=.7\linewidth]{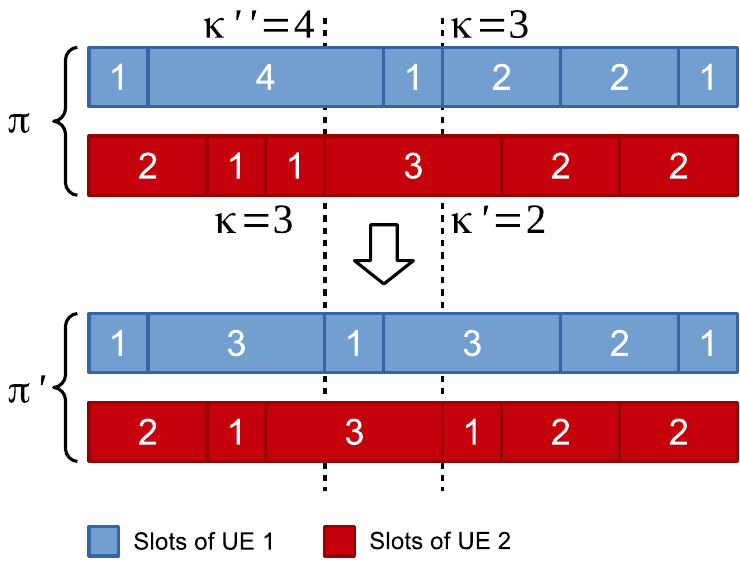}
			\subcaption{Inter-transmission intervals before and after}
		\end{subfigure}
		\begin{subfigure}[b]{\linewidth}
			\centering
			\includegraphics[width=.7\linewidth]{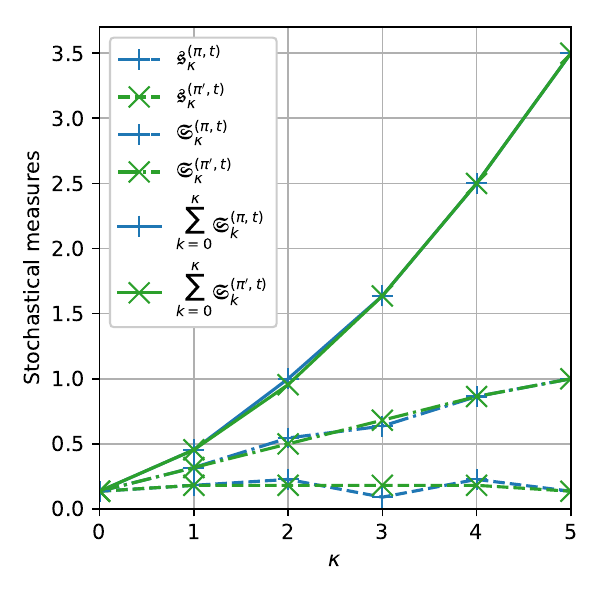}
			\subcaption{Second-order stochastic dominance achieved}
		\end{subfigure}
		\caption{Result of the slot exchange described in Tab.~\ref{tab:num_example_slot_exchange}}
		\label{fig:num_example_slot_exchange}
	\end{figure}
	
	For any given $t$, we can split the space of all schedules $\mathbf{\Pi}$ into two sets $\mathbf{\Pi}_1(t)$ and $\mathbf{\Pi}_2(t)$, where for all $\pi\in\mathbf{\Pi}_1(t)$, there exists $(x,y,\kappa,\kappa',\kappa'')$ to fulfill
	\begin{align}
	&\sum\limits_{j=1}^{k(\pi,y,t)-\kappa'-1}\tilde{s}_{y,j}^{(\pi,t)}<\sum\limits_{j=1}^{k(\pi,x,t)-\kappa}\tilde{s}_{x,j}^{(\pi,t)}<\sum\limits_{j=1}^{k(\pi,y,t)-\kappa'}\tilde{s}_{y,j}^{(\pi,t)},\label{eq:kappa_selection_1}\\
	&\sum\limits_{j=1}^{k(\pi,x,t)-\kappa''-1}\tilde{s}_{x,j}^{(\pi,t)}<\sum\limits_{j=1}^{k(\pi,y,t)-\kappa}\tilde{s}_{y,j}^{(\pi,t)}<\sum\limits_{j=1}^{k(\pi,x,t)-\kappa''}\tilde{s}_{x,j}^{(\pi,t)},\label{eq:kappa_selection_2}\\
	&\kappa'<\kappa<\kappa'' \label{eq:kappa_diff_constraint},
	\end{align}
	while $\nexists(x,y,\kappa,\kappa',\kappa'')$ for all $\pi\in\mathbf{\Pi}_2(t)$ to do so. Clearly, by recursively operating the slot assignment shifting process under the constraint \eqref{eq:kappa_diff_constraint}, any schedule $\pi\in\mathbf{\Pi}_1(t)$ can be eventually converted into some schedule $\pi'\in\mathbf{\Pi}_2(t)$, where for every single shifting step the stochastic dominance \eqref{eq:single_slot_assignment_swap_dominance} holds, so that $\sum\limits_{k=1}^\kappa\mathfrak{S}_{\pi',t}(\kappa)-\sum\limits_{k=1}^\kappa\mathfrak{S}_{\pi,t}(\kappa)<0$. Thus, we can assert that all schedules in $\mathbf{\Pi}_2(t)$ dominates all schedules in $\mathbf{\Pi}_1(t)$. 
	
	Moreover, by recursively operating the slot assignment shifting process under a modified constraint $\kappa''=\kappa=\kappa'+1$ instead of \eqref{eq:kappa_diff_constraint}, all schedules in $\mathbf{\Pi}_2(t)$ can be converted into each other after finite steps, with
	\begin{equation}
	\sum\limits_{k=0}^\kappa\mathfrak{S}^{(\pi',t)}_k-\sum\limits_{k=0}^\kappa\mathfrak{S}^{(\pi,t)}_k=0,\quad\forall(\pi,\pi')\in\mathbf{\Pi}_2^2(t).\label{eq:single_slot_assignment_shift_equivalence}
	\end{equation}
	This can be derived in the same way as we obtained \eqref{eq:single_slot_assignment_swap_dominance}, for which we omit the analytical proof here. A numerical example is provided in Tab.~\ref{tab:num_example_slot_exchange_equivalence} and Fig.~\ref{fig:num_example_slot_exchange_equivalence}.
	
	\begin{table}[!htbp]
		\centering
		\caption{An example of transmission slot exchange within $\mathbf{\Pi}_2(t)$}
		\label{tab:num_example_slot_exchange_equivalence}
		\begin{tabular}{l|cccccccccc}
			\toprule[2px]
			Transmission slot&1&2&3&4&5&6&7&8&9&10\\
			\midrule[1px]
			Scheduled UE ($\pi$)&2&1&1&2&\textbf{2}&\textbf{1}&1&2&2&1\\
			Scheduled UE ($\pi'$)&2&1&1&2&\textbf{1}&\textbf{2}&1&2&2&1\\
			\bottomrule[2px]
		\end{tabular}
	\end{table}
	\begin{figure}[!htbp]
		\centering
		\begin{subfigure}[b]{\linewidth}
			\centering
			\includegraphics[width=.7\linewidth]{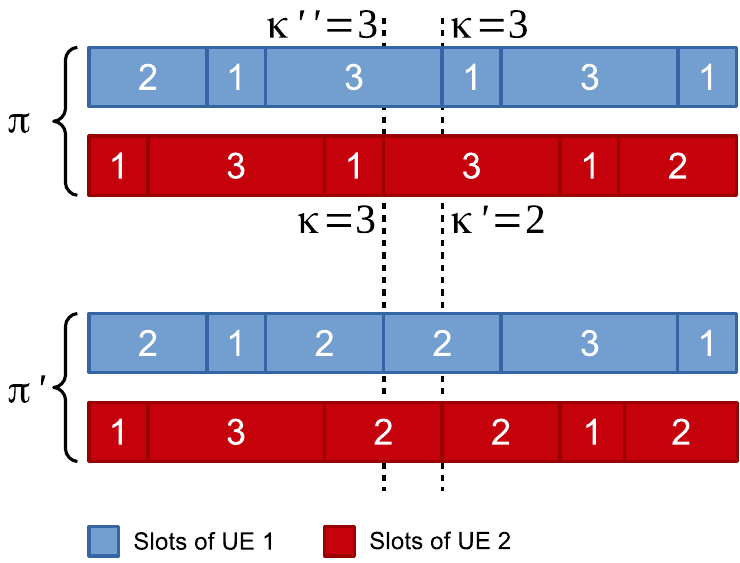}
			\subcaption{Inter-transmission intervals before and after}
		\end{subfigure}
		\begin{subfigure}[b]{\linewidth}
			\centering
			\includegraphics[width=.7\linewidth]{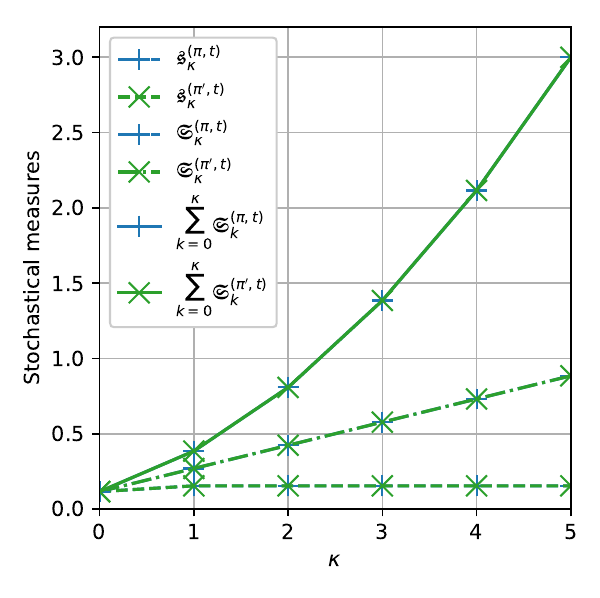}
			\subcaption{Second-order stochastic equivalence achieved}
		\end{subfigure}
		\caption{Result of the slot exchange described in Tab.~\ref{tab:num_example_slot_exchange_equivalence}}
		\label{fig:num_example_slot_exchange_equivalence}
	\end{figure}
	
	\emph{In summary, it holds that given a certain $t\in\mathbb{N}^+$, any schedule in $\mathbf{\Pi}_2(t)$ second-order stochastically dominates all valid schedules at $t$. Especially, round-robin is the only class of schedules that falls in $\mathbf{\Pi}_2(t)$ for all $t$, which makes it the unique class of schedules that dominate any other schedule at any time}.
	
	Furthermore, notice that $\varepsilon\subscript{\revise{avg}}^\kappa$ is a convex and monotonically decreasing function of $\kappa$, as proven in \cite{hanoch1969efficiency} we know that $\revise{\bar h\subscript{RR}^{(t,I)}}$ is statistically dominated by any other schedule.
	
	As the last step, now we consider an arbitrary initial state $\mathbf{S}^{(\pi,1)}$:
	\begin{equation}
		\mathbf{s}_{i,1}^{(\pi,1)}=\left({s}_{i,1}^{(\pi,1)}-1+1\right)\triangleq\left(\hat{{s}}_{i,1}^{(\pi,1)}+1\right),\quad\forall i\in\mathcal{I},
	\end{equation}
	where $\hat{\mathbf{s}}_i^{(\pi,1)}\geqslant 0$ for all $i\in\mathcal{I}$. Note that \eqref{eq:mean_aoi_under_schedule} still holds and can be now rewritten as
	\begin{align}
			&\bar\alpha^{(\pi,t)}=\frac{1}{I}\sum\limits_{i\in\mathcal{I}}\sum\limits_{\kappa=0}^{k(\pi,i,t)}\tilde{s}_{i,k(\pi,i,t)-\kappa}^{(\pi,t)}\varepsilon\subscript{\revise{avg}}^{\kappa}\nonumber\\
			=&\frac{1}{I}\sum\limits_{i\in\mathcal{I}}\left(\sum\limits_{\kappa=0}^{k(\pi,i,t)-2}s_{i,k(\pi,i,t)-\kappa}^{(\pi,t)}\varepsilon\subscript{\revise{avg}}^{\kappa}+s_{i,1}^{(\pi,1)}\varepsilon\subscript{\revise{avg}}^{k(\pi,i,t)-1}\right)\nonumber\\
			=&\underset{\bar{\alpha}_1}{\underbrace{\frac{1}{I}\sum\limits_{i\in\mathcal{I}}\left(\sum\limits_{\kappa=0}^{k(\pi,i,t)-2}s_{i,k(\pi,i,t)-\kappa}^{(\pi,t)}\varepsilon\subscript{\revise{avg}}^{\kappa}+\varepsilon\subscript{\revise{avg}}^{k(\pi,i,t)-1}\right)}}\nonumber\\
			&+\underset{\bar{\alpha}_2}{\underbrace{\frac{1}{I}\sum\limits_{i\in\mathcal{I}}\hat{s}_{i,1}^{(\pi,1)}\varepsilon\subscript{\revise{avg}}^{k(\pi,i,t)-1}}},
	\end{align}	
	where the first term $\bar{\alpha}_1$ is the average expected AoI with zero initial state and the second term $\bar{\alpha}_2$ is the impact of \revise{$\hat{s}_{i,1}^{(\pi,1)}$}. Clearly we can prove $\lim\limits_{t\to+\infty}\frac{\alpha_2}{\alpha_1}=0$, which implies that the impact of initial AoI state on the average expected AoI at $t$ horizon asymptotically converges to 0 as $t\to+\infty$. Thus, we come to Theorem~\ref{th:blind_optimum_rr}.
\end{proof}

\revise{\begin{proof}[Proof of Corollary~\ref{cr:blind_optimum_crr}]
	For the generic case where $l\in\mathcal{I}$, we slightly modify the transmission slot exchange operation defined to prove Theorem~\ref{th:blind_optimum_rr} as follows. In the first step, instead of finding a time slot $\tau_1$ for any given $(t,\pi,\kappa,x,y)$, we look for a tuple $(\tau_1,y)$ for any given $(t,\pi,\kappa,x)$, where $x$ is scheduled by $\pi$ to make its $(k(\pi,x,t)-\kappa)\superscript{th}$ transmission at $\tau_1$ while $y$ is idle at $\tau_1$. Thus, in the second step, we can update $\pi$ to $\pi''$ by replacing $x$ with $y$ in slot $\tau_1$. Furthermore, in the third step, instead of $\tau_2=\sum\limits_{j=1}^{k(x,y,t)-\kappa}\tilde{s}_{y,j}^{(\pi,t)}$, it shall select $\tau_2=\max\left\{\tau~\left\vert~\tau\leqslant\sum\limits_{j=1}^{k(x,y,t)-\kappa}\tilde{s}_{y,j}^{(\pi,t)},x\notin\mathcal{I}_{\tau}\right.\right\}$, and update $\pi''$ to $\pi'$ by replacing $y$ with $x$ in slot $\tau_2$. If no such $\tau_2$ exists, we take $\tau_2=0$ and $\pi'=\pi''$. 
	
	The remainder of the proof follows exactly that of Theorem~\ref{th:blind_optimum_rr}: by forcing $\kappa'\leqslant\kappa''-2$ where $\kappa''=\arg\limits_J\left\{\sum_{j=J}^{k(\pi',x,t)}\tilde{s}^{\pi',t}_{x,j}=t-\tau_2\right\}$ we can guarantee a stochastic dominance preferring $\pi'$ over $\pi$, so that it splits the complete blind schedule set $\mathbf{\Pi}$ into $\mathbf{\Pi}_1(t)$ and $\mathbf{\Pi}_2(t)$ for all $t$ according to Eqs.~\eqref{eq:kappa_selection_1}-\eqref{eq:kappa_diff_constraint}, where all schedules in $\mathbf{\Pi}_2(t)$ dominate any one in $\mathbf{\Pi}_1(t)$. Letting $\kappa''=\kappa=\kappa'+1$ we can also prove the stochastic equivalence among all schedules within $\mathbf{\Pi}_2(t)$ as we did when proving Theorem~\ref{th:blind_optimum_rr}, so that $\mathbf{\Pi}_2(t)$ second-order stochastically dominates all valid schedules at $t$. Since only $l$-clustered round-robin schedules are guaranteed to fall in $\mathbf{\Pi}_2(t)$ for all $t\in\mathbb{N}^+$, Corollary~\ref{cr:blind_optimum_crr} is proven.
\end{proof}}

\section{Proof of Theorem~\ref{th:l-opt}}\label{app:l-opt}
\begin{proof}
	Relaxing $l\in\mathbb{N}^+$ to $l\in\mathbb{R}^+$, it is trivial to derive that $\bar{h}_{\pi}^{(\infty,I)}$ is continuously differentiable about $l\in\mathbb{R}^+$. Thus, we can identify the extrema of $\bar{h}_{\pi}^{(\infty,I)}$ by forcing $\partial \bar{h}\subscript{RR}^{(\infty,I)}/\partial l\vert_{l=l\subscript{opt}}=0$. Since
	\begin{equation}
	\begin{split}
	&\frac{\partial\bar{h}_{\pi}^{(\infty,I)}}{\partial l}=\frac{\partial}{\partial l}\left(n\times \frac{1+\varepsilon\subscript{\revise{avg}}}{2(1-\varepsilon\subscript{\revise{avg}})}\times\frac{I}{l}\right)\\
	=&-\frac{nI}{l^2}\frac{1+\varepsilon\subscript{\revise{avg}}}{2(1-\varepsilon\subscript{\revise{avg}})}+\frac{nI}{l}\frac{\partial}{\partial l}\left(\frac{1+\varepsilon\subscript{\revise{avg}}}{2(1-\varepsilon\subscript{\revise{avg}})}\right)\\
	=&\frac{-nI(\varepsilon\subscript{\revise{avg}}+1)}{2l^2(1-\varepsilon\subscript{\revise{avg}})}+\frac{nI}{l(1-\varepsilon\subscript{\revise{avg}})^2}\frac{\partial\varepsilon\subscript{\revise{avg}}}{\partial l}\\
	=&\frac{-nI(\varepsilon\subscript{\revise{avg}}+1)}{2l^2(1-\varepsilon\subscript{\revise{avg}})}+\frac{nI(MC'n+\tau'l)}{2l^2(1-\varepsilon\subscript{\revise{avg}})^2\sqrt{2\pi MVne^{\beta^2}l}},
	\end{split}\label{eq:zero_forcing_l_opt}
	\end{equation}
	and $\frac{nI}{2l^2(1-\varepsilon\subscript{\revise{avg}})}>0$, it holds $l=l\subscript{opt}$ when
	\begin{equation}
	MC'n+\tau'l-\left(1-\varepsilon\subscript{\revise{avg}}^2\right)\sqrt{2\pi MVne^{\beta^2}l}=0.\label{eq:opt_l}
	\end{equation}
	Especially, In the FBL regime we generally consider $\varepsilon_i^2\approx 0$, so that recalling $\beta=\frac{MC'n/l-\tau'}{\sqrt{MVn/l}}$, from \eqref{eq:opt_l} we have
	\begin{align}
	\frac{MC'n+\tau'l\subscript{opt}}{MVnl\subscript{opt}}-2\pi e^{\frac{(MC'n-\tau'l\subscript{opt})^2}{MVnl\subscript{opt}}}&=0,&\\
	\frac{4C'\tau'}{V}+\beta\subscript{opt}^2-2\pi e^{\beta\subscript{opt}^2}&=0,&\label{eq:opt_l_approx}
	\end{align}
	where 
	\begin{equation}
	\beta\subscript{opt}\triangleq\beta\vert_{l=l\subscript{opt}}=\frac{MC'n-\tau'l\subscript{opt}}{\sqrt{MVnl\subscript{opt}}}.\label{eq:beta_opt}
	\end{equation}
	Let
	\begin{equation}
	w=-\left(\beta\subscript{opt}^2+\frac{4C'\tau'}{V}\right),\label{eq:w_def}
	\end{equation}
	Eq.~\eqref{eq:opt_l_approx} is reformed into:
	\begin{equation}
	we^w=-2\pi e^{-\frac{4C'\tau'}{V}}.\label{eq:lambert_equation}
	\end{equation}
	
	By solving Eq.~\eqref{eq:lambert_equation} (detailed in Appendix~\ref{app:optimal_cluster_size}) we are able to prove that the only extremum of $\bar{h}_\pi^{(\infty,I)}$ is located at $l\subscript{opt}$ provided by Theorem~\ref{th:l-opt}. Then we further observe the second derivative:
	\begin{align}
	&\frac{\partial^2\bar{h}_{\pi}^{(\infty,I)}}{\partial l^2}=\frac{nI}{2}\frac{\partial}{\partial l}\left[\underbrace{\frac{1}{l^2\left(1-\varepsilon\subscript{\revise{avg}}\right)^2}}_{X_1}\underbrace{\left(\frac{C'Mn+\tau'l}{\sqrt{2\pi MVnle^{\beta^2}}}-1\right)}_{X_2}\right]\nonumber\\
	=&\frac{nI}{2}\left(X_1\frac{\partial X_2}{\partial l}+X_2\frac{\partial X_1}{\partial l}\right)
	\end{align}
	
	Especially, when $l=l\subscript{opt}$, according to Eq.~\eqref{eq:beta_opt}, $X_2=0$, $\sqrt{2\pi MVnle^{\beta^2}}=C'Mn+\tau'l$, and
	\begin{equation}
		\begin{split}
			&\frac{\partial X_2}{\partial l}=\frac{\partial}{\partial l}\left(\frac{C'Mn+\tau'l}{\sqrt{2\pi MVnl}}\times\frac{1}{\sqrt{e^{\beta^2}}}\right)\\
			=&\frac{C'Mn+\tau'l}{\sqrt{2\pi MVnl}}\times\frac{-\beta}{\sqrt{e^{\beta^2}}}\times\frac{-\left(C'Mn+\tau'l\right)}{2l\sqrt{MVnl}}\\
			&+\frac{1}{\sqrt{e^{\beta^2}}}\times\frac{-\left(C'Mn-\tau'l\right)}{2l\sqrt{2\pi MVn}}\\
			=&\frac{C'Mn-\tau'l}{2l\left(C'Mn+\tau'l\right)MVnl}\left[\left(C'Mn+\tau'l\right)^2-MVnl\right]
		\end{split}
	\end{equation}
	
	It is trivial to derive $X_1>0$, $C'Mn-\tau'l\subscript{opt}>0$, and $2C'\tau'>V$. Therewith we assert $\frac{\partial^2\bar{h}\subscript{RR}^{\infty,I}}{\partial l^2}>0$ at $l=l\subscript{opt}$, \revise{so that Theorem~\ref{th:l-opt} guarantees to provide the global minimum}.
\end{proof}

\section{Proof of Lemma~\ref{lemma:blind assignment}}\label{app:blind_assignment}
\begin{proof}
	Firstly, we investigate the convexity of $\varepsilon_{i}$. 
	The second derivative of $\varepsilon_i$ with respect to $0\leqslant b_{i} \leqslant M$ is given by:
	\begin{equation}
		\begin{split}
			\frac{{{\partial ^2}{\varepsilon _{i}}}}{{\partial b_{i}^2}} =&\int\left[\beta_{i}\exp\left(-\frac{{\beta _{i}^2}}{2}\right)\left( {\beta_{i} {{\left( {\frac{{\partial {\beta _{i}}}}{{\partial {b_{i}}}}} \right)}^2} - \frac{{{\partial ^2}{\beta _{i}}}}{{\partial b_{i}^2}}} \right)\right.\\&\left.\times\prod\limits_{m\in\mathcal{M}} f_Z(z_{m,i})\right]\diff\mathbf{Z}
		\end{split}
	\end{equation}
	The derivative is non-negative if both $-\frac{{{\partial ^2}{\beta _{i}}}}{{\partial b_{i}^2}}$ and $\beta_{i}$ are non-negative. For the first term, We have:
	\begin{equation}
	\label{eq:error_second}
	\frac{{{\partial ^2}{\beta _{i}}}}{{\partial b_{i}^2}} =  - \frac{{\left( {{b_{i}}\sum\limits_{m\in\mathcal{M}} {{C_{m,i}}n + 3\tau } } \right)}}{{4b_{i}^2n^2\sum\limits_{m\in\mathcal{M}} { {{V_{m,i}}}} }}
	\end{equation}
	Since both the nominator and denominator are non-negative, it holds that $-\frac{{{\partial ^2}{\beta _{i}}}}{{\partial b_{i}^2}}\geqslant 0$. 
	However, the sign of $\beta_{i}$ depends on the channel gain $z_{m,i}$. Therefore, for any given $\mathbf{Z}$, $\varepsilon_{i}$ is convex(concave) if the instantaneous Shannon capacity is greater(smaller) than the coding rate. In particular, let $0\leqslant \mathbf{\xi}<\mathbf{\hat{z}}$ denote the channel gain threshold so that the coding rate is always lower than the instantaneous Shannon capacity if the channel gain is higher than $\xi$. Then, $\varepsilon_{i}$ can be expressed as the sum of three integral parts, i.e. \revise{$\frac{{{\partial ^2}{\beta _{i}}}}{{\partial b_{i}^2}} =\underbrace{\int^{\mathbf{\xi}}_{\mathbf{0}}\Delta({\mathbf{Z}})\diff\mathbf{Z}}_{<0}+\underbrace{\int^{\mathbf{\hat{z}}}_{\mathbf{\xi}}\Delta({\mathbf{Z}})\diff\mathbf{Z}}_{>0}+\underbrace{\int^{\mathbf{\infty}}_{\mathbf{\hat{z}}}\Delta({\mathbf{Z}})\diff\mathbf{Z}}_{>0}$},
	where $\Delta(\mathbf{Z})=\beta_{i} \exp \left( { - \frac{{\beta _{i}^2}}{2}} \right)\left( {\beta_{i} {{\left( {\frac{{\partial {\beta _{i}}}}{{\partial {a_{\pi,m,i}\revise{(t)}}}}} \right)}^2} - \frac{{{\partial ^2}{\beta _{i}}}}{{\partial a_{\pi,m,i}^2\revise{(t)}}}} \right) \prod\limits_{m\in\mathcal{M}} f_Z(z_{m,i})$. Since $\mathbf{\hat{z}}$ is the median, it holds $\int^{\mathbf{\hat{z}}}_{\mathbf{0}}\prod f_Z(z_{m,i})d{\mathbf{Z}}=\int^{\infty}_{\mathbf{\hat{z}}}\prod f_Z(z_{m,i})d{\mathbf{Z}}=\frac{1}{2}\geqslant \int^{\mathbf{\xi}}_{\mathbf{0}}\prod f_Z(z_{m,i})d{\mathbf{Z}}$. Furthermore, we have:
	\begin{equation}
	\begin{split}
	    &\int^\infty_{\mathbf{\xi}} \beta_{i} e^{ - \frac{{\beta _{i}^2}}{2}}\left( {\beta_{i} {{\left( {\frac{{\partial {\beta _{i}}}}{{\partial {a_{\pi,m,i}\revise{(t)}}}}} \right)}^2} - \frac{{{\partial ^2}{\beta _{i}}}}{{\partial a_{\pi,m,i}^2\revise{(t)}}}} \right)\diff\mathbf{Z}\\
	    \geqslant&\left\vert\int^{\mathbf{\xi}}_{\mathbf{0}} \beta_{i} e^{ - \frac{{\beta _{i}^2}}{2}}\left( {\beta_{i} {{\left( {\frac{{\partial {\beta _{i}}}}{{\partial {a_{\pi,m,i}\revise{(t)}}}}} \right)}^2} - \frac{{{\partial ^2}{\beta _{i}}}}{{\partial a_{\pi,m,i}^2\revise{(t)}}}} \right)\diff\mathbf{Z}\right\vert
	\end{split}
	\end{equation}
    Hence, we have
	\revise{$\int^{\mathbf{\infty}}_{\mathbf{\hat{z}}}\Delta({\mathbf{Z}})\diff\mathbf{Z}\geqslant \left|\int^{\mathbf{\xi}}_{\mathbf{0}}\Delta({\mathbf{Z}})\diff\mathbf{Z} \right|$. Especially, it takes the equality only when} $\xi=\hat{z}$.
	As a result, we have $\frac{{{\partial ^2}{\varepsilon _{i}}}}{{\partial b_{i}^2}} \geqslant 0$, i.e., $\varepsilon_{i}$ is convex in $b_{i}$.
Therefore, $\varepsilon_i$ is convex in the relaxation of $\mathbf{b}$.
	Next, let $t \geqslant \varepsilon_{i}$. We formulate the partial Lagrangian for Problem~\eqref{problem_blocklength_blind} as
	$L=t+\sum \lambda_i(\varepsilon_{i}-t)+\mu\left(\sum\limits_{m \in \mathcal{M}} b_{i}-1\right)$.
	Based on the KKT conditions, \revise{for all $i\in\mathcal{I}$ the  optimum $(t,\lambda_i,\nu)$ must satisfy the necessary  and sufficient conditions of $\lambda^*_i \geqslant 0$, $\lambda^*_i(\varepsilon^*_i-t^*)=0$, and $\frac{\partial L}{\partial b_{i}}=\frac{\partial \varepsilon_i}{\partial b_{i}}+\mu^*=0$.}
	Note that $b_{i}$ is finite, i.e., $\frac{\partial \varepsilon_i}{\partial b_{i}}<0$ always holds. Therefore, we have 	
	$\frac{\partial \varepsilon_{1}}{\partial b_{1}}=\frac{\partial \varepsilon_{2}}{\partial b_{2}}=...=\mu$.
	Since $\varepsilon_{i}$ is a convex and monotonically decreasing function, the equality chain holds only when
	\begin{equation}
	\varepsilon_1=\varepsilon_2=...=\tilde{\mu}.
    \label{eq:eps_uniform}
	\end{equation}
	Hence, the optimal sub-carrier allocation that satisfies~\eqref{eq:eps_uniform} is $b_{1}=b_{2}=...=\frac{M}{l}$.
\end{proof}

\section{Solving the Optimal UE Cluster Size}\label{app:optimal_cluster_size}
In Eq.~\eqref{eq:lambert_equation}, since $w\in\mathbb{R}$, it takes two branches of the Lambert $W$ function: $w=W_0\left(-2\pi e^{-\frac{4C'\tau'}{V}}\right)$ and $w=W_{-1}\left(-2\pi e^{-\frac{4C'\tau'}{V}}\right)$, each generating two roots of the equation
\begin{equation}
	w=-\left(\beta^2+\frac{4C'\tau'}{V}\right)=-\frac{\left(MC'n-\tau'l\subscript{opt}\right)^2}{MVnl\subscript{opt}}-\frac{4C'\tau'}{V}.\label{eq:quadric_eq_l_opt}
\end{equation}
Thus, up to four roots $l\subscript{opt}$ can be obtained. However, as discussed in Sec.~\ref{subsec:rr_optimum}, the convexity of $\bar{h}_\pi^{(\infty,I)}$ ensures an \emph{unique} root $l\subscript{opt}\in[1,M]$, so three false roots shall be rejected for our problem.

First consider the principal branch $w=W_0\left(-2\pi e^{-\frac{4C'\tau'}{V}}\right)$, which generates two roots $l_{\text{opt},1}$ and $l_{\text{opt},2}$. Since it generally holds $4C'\tau'\gg V$, we have $-2\pi e^{-\frac{4C'\tau'}{V}}\approx 0$. It is known that $W_0(x)\approx x$ when $x\approx 0$, so 
\revise{\eqref{eq:quadric_eq_l_opt} suggests that for all $i\in\{1,2\}$,
$-\frac{\left(MC'n-\tau'l_{\text{opt},i}\right)^2}{MVnl_{\text{opt},i}}-\frac{4C'\tau'}{V}\approx 0
	\Rightarrow\left(MC'n+\tau'l_{\text{opt},i}\right)^2\approx 0$, therefore $l_{\text{opt},1}\approx l_{\text{opt},2}\approx-\frac{MC'n}{\tau'}<0$.}
Since $l\subscript{opt}\in[1,M]$, both $l_{\text{opt},1}$ and $l_{\text{opt},2}$ are rejected.

Then we investigate the lower branch $	w=W_{-1}(-2\pi e^{-\frac{4C'\tau'}{V}})<-\frac{4C'\tau'}{V}$, which returns two positive real roots of \eqref{eq:quadric_eq_l_opt},
\revise{namely $l\subscript{opt,3}=\frac{1}{2\tau'^2}\times\left(\delta+\sqrt{\delta^2-4\tau'^2M^2C'^2n^2}\right)$ and $l\subscript{opt,4}=\frac{1}{2\tau'^2}\times\left(\delta-\sqrt{\delta^2-4\tau'^2M^2C'^2n^2}\right)$.}
Here, remark that \eqref{eq:quadric_eq_l_opt} is derived from \eqref{eq:zero_forcing_l_opt} under the approximation $\varepsilon\subscript{\revise{avg}}^2\approx 0$. This only holds when $\beta>0$ is significant, which requires
\begin{equation}
	l<\frac{MC'n}{\tau'}=\frac{MCn}{\tau}.\label{eq:l_ub}
\end{equation}
Recalling  \eqref{eq:delta}, \eqref{eq:beta_opt}, and \eqref{eq:w_def}, we have
\begin{align}
	&\delta=2MC'n\tau'+MVn\beta\subscript{opt}^2=\frac{M^2C'^2n^2+\tau'^2l\subscript{opt}^2}{l\subscript{opt}},\\
	&\sqrt{\delta^2-4\tau'^2M^2C'^2n^2}=\frac{M^2C'^2n^2-\tau'^2l\subscript{opt}^2}{l\subscript{opt}}.
\end{align}
Thus, if $l\subscript{opt}=l\subscript{opt,3}$, 
\revise{we have} $l\subscript{opt,3}=\frac{M^2C'^2n^2}{\tau'^2l_{\text{opt},3}}$,
which gives $
	l_{\text{opt},3}=\frac{MC'n}{\tau'}=\frac{MCn}{\tau}
$ that violates the constraint \eqref{eq:l_ub}, so that $l_{\text{opt},3}$ is rejected. To the contrary, for $l\subscript{opt,4}$ we have \revise{$l\subscript{opt,4}=\frac{\delta-\sqrt{\delta^2-4\tau'^2M^2C'^2n^2}}{2\tau'^2}<\frac{\sqrt{\delta^2-\left(\delta^2-4\tau'^2M^2C'^2n^2\right)}}{2\tau'^2}=\frac{MC'n}{\tau'}=\frac{MCn}{\tau},$}
which satisfies \eqref{eq:l_ub} and therewith ensures $
\beta\vert_{l=l_{\text{opt},4}}>0\Rightarrow\varepsilon\subscript{\revise{avg}}\vert_{l=l_{\text{opt},4}}<0.5.
$ Thus, $l_{\text{opt},4}$ is confirmed as the only capable optimum of cluster size $l\subscript{opt}$ that minimizes $\bar{h}_{i,\pi}^{\infty,I}$.


\ifCLASSOPTIONcaptionsoff
  \newpage
\fi


%
%

\vfill

\begin{IEEEbiography}[{\includegraphics[width=1in,height=1.25in,clip,keepaspectratio]{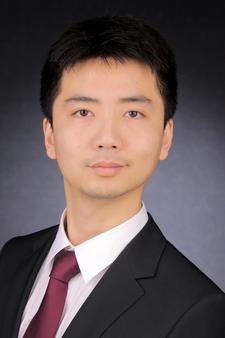}}]{Bin Han} (M'15--SM'21)
received in 2009 his B.E. degree from Shanghai Jiao Tong University, M.Sc. in 2012 from Technical University of Darmstadt, and in 2016 the Ph.D. degree from Karlsruhe Institute of Technology. Since July 2016 he has been with University of Kaiserslautern as Postdoctoral Researcher and Senior Lecturer, researching in the broad area of wireless communication and networking. He is the author of over 40 papers and book chapters, participated in multiple EU collaborative research projects, and serves as TPC member of GLOBECOM, EuCNC, and European Wireless.
\end{IEEEbiography}

\vfill

\begin{IEEEbiography}[{\includegraphics[width=1in,height=1.25in,clip,keepaspectratio]{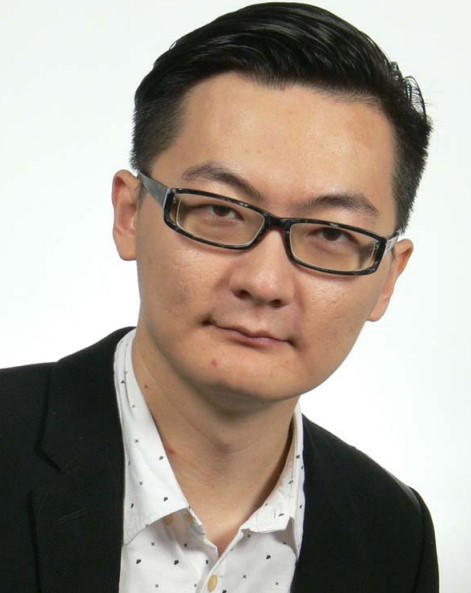}}]{Yao Zhu} (S'19) received the B.S. degree in electrical engineering from the University of Bremen, Bremen, Germany, in 2015, and the master’s degree in information technology and computer engineering from RWTH Aachen University, Aachen, Germany, in 2018. He is currently pursuing the Ph.D. degree with the ISEK Research Group, RWTH Aachen University. His research interests include ultra-reliable and low-latency communications, and mobile edge networks.
\end{IEEEbiography}

\vfill

\begin{IEEEbiography}[{\includegraphics[width=1in,height=1.25in,clip,keepaspectratio]{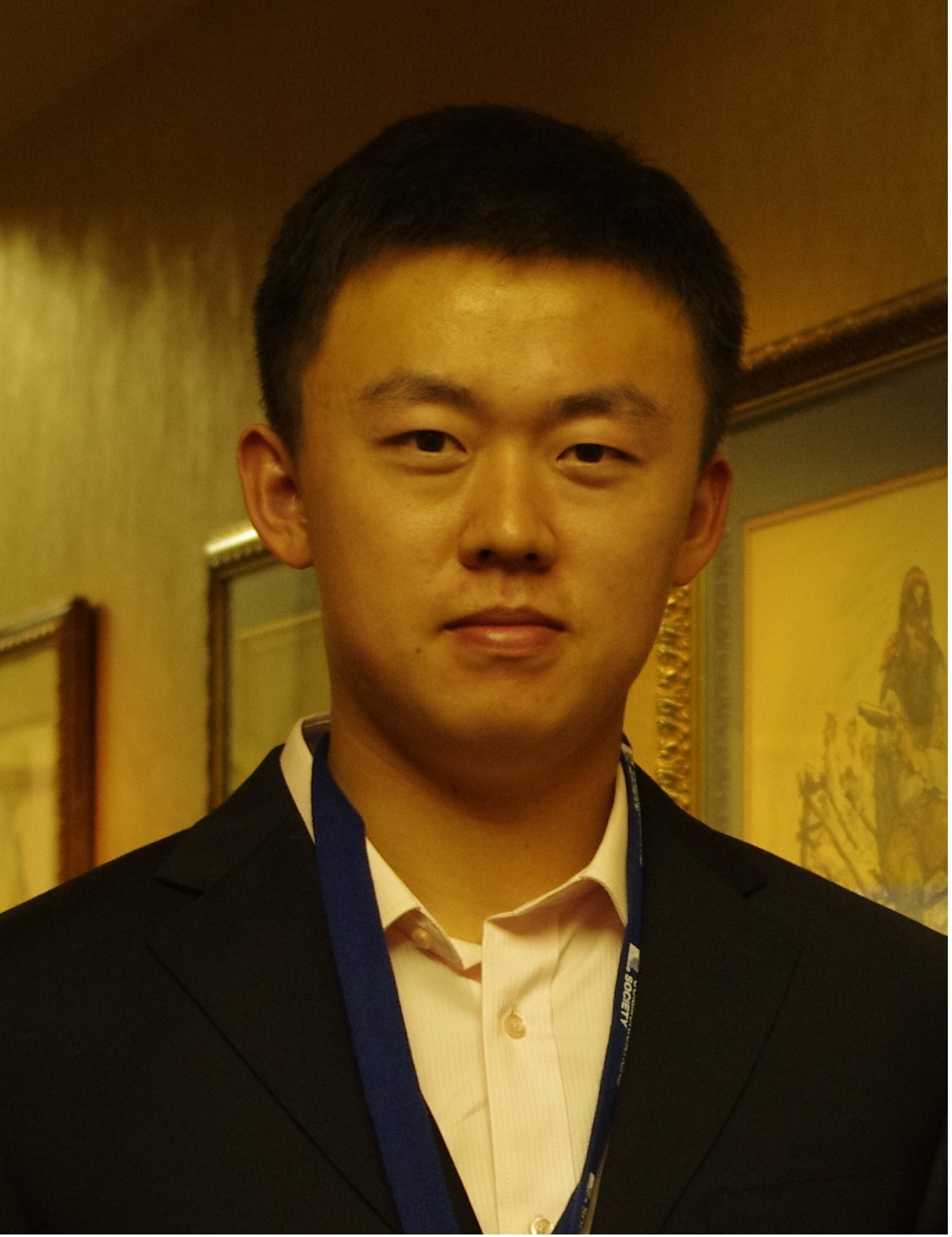}}]{Zhiyuan Jiang} (M'15)
received the B.S. and Ph.D. degrees from the Department of Electronic Engineering, Tsinghua University, China, in 2010 and 2015, respectively. He visited the WiDeS Group, University of Southern California, Los Angeles, CA, USA, from 2013 to 2014. He is currently a professor with the School of Communication and Information Engineering, Shanghai University, China. He serves as a TPC Member for IEEE INFOCOM, ICC, GLOBECOM, and WCNC. He serves as an Associated Editor for IEEE/ KICS Journal of Communications and Networks and a Guest Editor for IEEE Internet of Things Journal. He received the ITC Rising Scholar Award in 2020, the Best Paper Award from the IEEE ICC 2020, the Best in-Session Presentation Award from the IEEE INFOCOM 2019, and an Exemplary Reviewer Award from IEEE WCL in 2019. His current research interests include URLLC in wireless networked control systems and V2X systems.
\end{IEEEbiography}

\vfill

\begin{IEEEbiography}[{\includegraphics[width=1in,height=1.25in,clip,keepaspectratio]{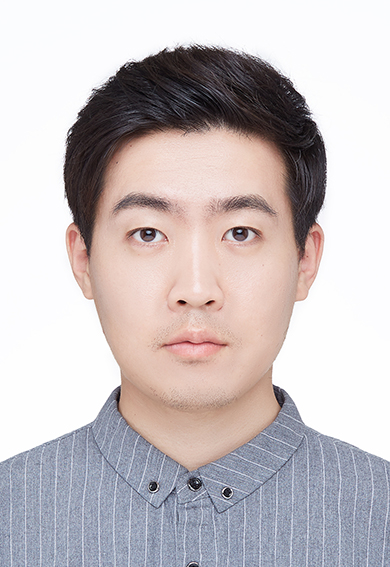}}]{Muxia Sun}
		received in 2010 his B.Sc. degree from South China University of Technology (SCUT), M.Sc. in 2012 \& 2013 from Universit\'e de Nantes and SCUT, respectively, and the Ph.D. degree in 2019 from Universit\'e Paris-Saclay. Since 2020 he has been with Tsinghua University as Postdoctoral Researcher in the Department of Industrial Engineering. His current research interests include reliability assessment and optimization of industrial \& communication systems, robust optimization, and approximation algorithm design.
\end{IEEEbiography}


\begin{IEEEbiography}[{\includegraphics[width=1in,height=1.25in,clip,keepaspectratio]{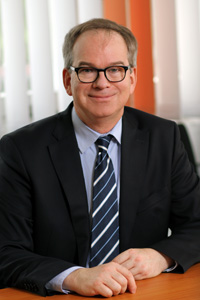}}]{Hans D. Schotten} (S'93--M'97) received the Ph.D. degrees from the RWTH Aachen University of Technology, Germany, in 1997. From 1999 to 2003, he worked for Ericsson. From 2003 to 2007, he worked for Qualcomm. He became manager of a R\&D group, Research Coordinator for Qualcomm Europe, and Director for Technical Standards. In 2007, he accepted the offer to become the full professor at the University of Kaiserslautern. In 2012, he - in addition - became scientific director of the German Research Center for Artificial Intelligence (DFKI) and head of the department for Intelligent Networks. Professor Schotten served as dean of the department of Electrical Engineering of the University of Kaiserslautern from 2013 until 2017. Since 2018, he is chairman of the German Society for Information Technology and member of the Supervisory Board of the VDE. He is the author of more than 200 papers and participated in 40+ European and national collaborative research projects.
\end{IEEEbiography}

\vfill

%
%
%




\clearpage

\end{document}